\documentclass[pra,twocolumn,superscriptaddress,showpacs,amsmath,amstex,amssymb,citeautoscript]{revtex4-2}

\usepackage[T1]{fontenc}
\usepackage[utf8]{inputenc}

\usepackage{lipsum}
\usepackage{amsmath}
\usepackage{amssymb}
\usepackage{bbm}
\usepackage{braket}
\usepackage{xcolor}
\usepackage{pifont}
\usepackage[mathscr]{euscript}
\usepackage[shortlabels]{enumitem}

\usepackage{graphicx}
\usepackage{lipsum}
\allowdisplaybreaks
\usepackage{float}
\usepackage{graphicx}
\usepackage{dsfont}
\usepackage{comment}
\usepackage[colorlinks=true]{hyperref}  
\hypersetup{
    bookmarks=true,         
    unicode=false,          
    pdftoolbar=true,        
    pdfmenubar=true,        
    pdffitwindow=false,     
    pdfstartview={FitH},    
    pdftitle={Quantum geometry and impurity sensitivity of superconductors},    
    pdfauthor={Sedov and Scheurer},     
    pdfsubject={},   
    pdfcreator={},   
    pdfproducer={}, 
    pdfkeywords={} {} {}, 
    pdfnewwindow=true,      
    colorlinks=true,       
    linkcolor=blue, 
    citecolor=blue,        
    filecolor=magenta,      
    urlcolor=blue           
}

\newcommand{\equref}[1]{Eq.~(\ref{#1})}

\newcommand{\figref}[1]{Fig.~\ref{#1}}
\newcommand{\refcite}[1]{Ref.~\onlinecite{#1}}

\newcommand{\appref}[1]{Appendix~\ref{#1}}

\newcommand{\pdagger}{{\phantom{\dagger}}}

\newcommand{\sign}{\,\text{sign}}

\renewcommand{\vec}[1]{\mathbf{#1}}

\definecolor{wrongultramarine}{rgb}{1,0.5,0}

\renewcommand{\sign}{\mathrm{sign}}
\newcommand{\pd}{\phantom{\dagger}}

\usepackage{mathtools}
\usepackage[english]{babel}

\begin{document}

\title{Quantum geometry and impurity sensitivity of superconductors without time-reversal symmetry: application to rhombohedral graphene and altermagnets}

\author{Denis Sedov}
\affiliation{Institute for Theoretical Physics III, University of Stuttgart, 70550 Stuttgart, Germany}
\author{Mathias S.~Scheurer}
\affiliation{Institute for Theoretical Physics III, University of Stuttgart, 70550 Stuttgart, Germany}

\begin{abstract}
Analyzing the consequences of the quantum geometry induced by the momentum dependence of Bloch states has emerged as a very rich and active field in condensed matter physics. For instance, for the superfluid stiffness or the pairing mechanism, these geometric aspects can play an important role. We here demonstrate that quantum geometry can also be essential for the disorder sensitivity of a superconductor, in particular when time-reversal symmetry is broken in the normal-state Bloch Hamiltonian. We derive a general expression for the behavior of the critical temperature $T_c$ involving weighted (anti-)commutators of the superconducting order parameter and impurity matrix elements, which includes both wave-function effects and kinetic pair breaking due to broken time-reversal symmetry in the dispersion. 
We analyze how the former effects lead to ``quantum geometric pair breaking’’, where any superconductor becomes susceptible to microscopically non-magnetic impurities, and formally relate it to the maximum possible localization of two-particle Wannier states.
Further, in the presence of kinetic pair breaking, impurities can also enhance pairing, leading to an overall more complex, non-monotonic behavior of $T_c$ with impurity concentration. We also analyze the fate of finite-momentum pairing. 
Our results are directly relevant to rhombohedral graphene, twisted MoTe$_2$, and superconducting altermagnets. 
\end{abstract}

\maketitle

\section{Introduction}
Among all symmetries of a metallic system, time-reversal symmetry (TRS) plays a particularly central role in the formation and theoretical description of both conventional and unconventional superconductivity \cite{RevModPhys.63.239,Ghosh_2021,AndersonPaper}. It leads to a degeneracy of the band energies $\xi_{\vec{k}} = \xi_{-\vec{k}}$ at momenta $\vec{k}$ and $-\vec{k}$, which is the reason why Cooper pairs typically have zero center of mass momentum and what makes superconductivity favorable already at weak coupling and, hence, ubiquitous in metals at sufficiently low temperatures. This is also why recent indications of superconductivity emerging from a valley-imbalanced normal state in rhombohedral tetralayer graphene (R4G) \cite{rtg_chiral_sc_experiment} and twisted MoTe$_2$ \cite{MoTe2_unconv_exp} have attracted significant attention, see, e.g., \cite{PhysRevB.111.174523,LiangFuRhombohedral,YahuisPaper,JahinLin,QiongWu,YoonTightBindingDiode,StandfordWithTheRelativeChirality,Paco,zgnk-rw1p,TheSTMPaper,ChristosEnergetics,k8sb-rqxf,2025arXiv250419321G,FranzSCRingOfFire} and many more. 
An energetic shift between the two valleys breaks TRS in the orbital channel, which leads to a variety of interesting consequences for the superconductor, such as the natural emergence of topological superconductivity \cite{PhysRevB.111.174523,LiangFuRhombohedral,YahuisPaper,JahinLin,QiongWu,YoonTightBindingDiode,StandfordWithTheRelativeChirality,Paco,zgnk-rw1p,TheSTMPaper,ChristosEnergetics,k8sb-rqxf,2025arXiv250419321G,FranzSCRingOfFire}, a superconducting diode effect \cite{scammell_theory_2022,YoonTightBindingDiode,zgnk-rw1p}, translational-symmetry breaking pairing \cite{TheSTMPaper,ChristosEnergetics,k8sb-rqxf,2025arXiv250419321G}, and exotic tunneling signatures \cite{TheSTMPaper} to mention a few.

A second reason why TRS is central to the theoretical description of pairing is its role in gauge fixing. Consider a superconducting order parameter $\Delta_{\vec{k}}$ in a single (possibly spin-degenerate) band with Bloch states $\ket{\phi_{\vec{k}}}$. Under the gauge transformation 
\begin{equation}
\ket{\phi_{\vec{k}}} \rightarrow e^{i \alpha_{\vec{k}}} \ket{\phi_{\vec{k}}}, \label{gaugetrafro}
\end{equation}
the superconducting order parameter is not invariant but changes as $\Delta_{\vec{k}} \rightarrow e^{-i (\alpha_{\vec{k}}+\alpha_{-\vec{k}})} \Delta_{\vec{k}} $. Assuming for concreteness that spin-orbit coupling can be neglected (see \cite{DesignPrinciples} for the related discussion with spin-orbit coupling) and the system exhibits a spinless TRS with anti-unitary operator $\Theta$ (obeying $\Theta^2 = \mathbbm{1}$), we can fix the \textit{relative} phase at $\vec{k}$ and $-\vec{k}$ by choosing $\ket{\phi_{-\vec{k}}}=\Theta \ket{\phi_{\vec{k}}}$. Then $\alpha_{\vec{k}}$ are required to be odd in momentum and $\Delta_{\vec{k}}$ is invariant under \equref{gaugetrafro}. 
This is also the gauge in which the critical temperature of a $\vec{k}$-independent superconducting order parameter is unaffected by non-magnetic disorder as long as $k_F l \ll 1$ (with Fermi momentum $k_F$ and mean-free path $l$). This observation, which is known as Anderson’s theorem \cite{AndersonPaper,abrikosov1959theory,abrikosov1959superconducting}, can be intuitively thought of as a consequence of the preserved TRS \cite{AndersonPaper} and constitutes a third reason why TRS is crucial for superconductivity. 

Motivated by the aforementioned experiments on rhombohedral graphene and twisted MoTe$_2$ as well as the growing interest in the interplay of superconductivity and altermagnetism \cite{PhysRevX.12.040501,2025APPSB..35...18M,2025JPCM...37E3003F,2025arXiv251009170L,PhysRevB.108.054510,PhysRevB.111.L100502,f6nc-vsnx,PhysRevB.111.174436}, we here develop a general theory for pair breaking in systems with broken TRS in the normal state. Importantly, we not only take into account kinetic effects related to $\xi_{\vec{k}} \neq \xi_{-\vec{k}}$ but also the ``quantum geometry'' encoded in the $\vec{k}$-dependence of the Bloch states $\ket{\phi_{\vec{k}}}$. While the consequences of this form of quantum geometry have been very actively studied \cite{ReviewQGBernevig} for a variety of phenomena, e.g., for transport \cite{2024arXiv241219056C,PhysRevResearch.4.013001,PhysRevB.110.174423}, the superfluid stiffness \cite{ReviewQuantumGeometry,DisorderSFW,2025PhRvB.112i4501C,2025arXiv251005224K}, the finite-momentum superconducting response \cite{PhysRevB.108.094508}, pair size \cite{PhysRevB.111.014502}, spectral weight transfer \cite{PhysRevB.104.L100501}, Andreev bound-state localization \cite{2025arXiv250515891L}, and for pairing mechanisms \cite{PhysRevLett.134.176001,JahinLin,QGDiscrepancy,2025arXiv250913407S,StandfordWithTheRelativeChirality,ChristosEnergetics}, we here show that it also plays an important role in the disorder sensitivity of superconductivity, in particular, when TRS is broken: 
without any anti-unitary relation between $\ket{\phi_{\vec{k}}}$ and $\ket{\phi_{-\vec{k}}}$, nominally non-magnetic impurities are shown to be generically pair breaking for any superconductor as a result of the TRS-breaking contributions to quantum geometry. We define a measure of the degree of quantum geometric pair breaking, which is formulated as an optimization problem. As a byproduct, this allows us to fix the gauge in a natural way, even when TRS is absent, and is shown to be equivalent to finding the maximally localized (see definition below) Cooper pair wave function.
We also discuss the non-trivial additional consequences resulting from broken TRS in the dispersion and finite momentum pairing.
Our general results are illustrated using concrete models, relevant to the aforementioned systems, as examples.

\section{Results}
In this work, we are interested in the stability of superconductivity in the presence of impurities modeled by 
\begin{align}
    H_{\mathrm{dis}} = \sum_{\alpha,\beta}\int d \mathbf{r} \, V(\mathbf{r}) \psi^\dagger_\alpha(\mathbf{r}) w_{\alpha\beta} \psi_{\beta}^\pdagger(\mathbf{r}), \label{DisorderHamiltonian}
\end{align}
where $\psi^\dagger_{\alpha}(\mathbf{r})$ are the field operators creating an electron with internal index $\alpha$ (e.g., sublattice or layer quantum numbers) at position $\mathbf{r}$ (either in the continuum or on a lattice); $V(\mathbf{r})$ is the local quenched disorder potential, which we assume to be $\delta$-correlated, $\braket{V(\mathbf{r}) V(\mathbf{r}')}_{\mathrm{dis}} = \gamma^2 \mathcal{V} \delta(\mathbf{r} - \mathbf{r}')$ or $\braket{V_{\mathbf{q}} V_{-\mathbf{q}'}}_{\mathrm{dis}} = \gamma^2 \delta_{\mathbf{q},\mathbf{q}'}$ in Fourier space, with system volume $\mathcal{V}$ and disorder strength $\gamma$. Furthermore, $w$ is a matrix describing the internal structure of the impurities, in general only constrained by Hermiticity to obey $w^\dagger = w$. 

As superconductivity is typically a low-energy phenomenon, it is natural to go to the eigenbasis of the normal-state Bloch Hamiltonian $h_{\vec{k}}$ and further project onto the ``active band'' at the Fermi level with energies $\xi_{\vec{k}}$ and wave functions $\ket{\phi_{\vec{k}}}$, respectively. This is a common approximation in this context as superconducting matrix elements to higher-energy bands away from the Fermi level are often negligible, and since scattering processes involving other bands are suppressed by a factor of the squared inverse bandgap. On the level of field operators $\psi(\mathbf{r}) \to 1/\mathcal{V} \sum_{\mathbf{k}} \phi_{\mathbf{k}} e^{i\mathbf{kr}} d_{\mathbf{k}}$ and the impurity Hamiltonian becomes
\begin{align}
    H_{\mathrm{dis}} \to \sum_{\mathbf{k},\mathbf{k}'} V_{\mathbf{k}-\mathbf{k}'} d^\dagger_{\mathbf{k}} W_{\mathbf{k},\mathbf{k}'}  d^{\pd}_{\mathbf{k}'}, \, W_{\mathbf{k},\mathbf{k}'} =\braket{\phi_{\mathbf{k}}| w | \phi_{\mathbf{k}'}}.
\end{align} 
Our general analysis will capture three different cases concerning the spin degree of freedom: (1) if the normal state is already spin polarized (e.g., by a Zeeman field or the spontaneous formation of a ``quarter metal'' in rhombohedral graphene \cite{rtg_chiral_sc_experiment} or, locally in $\vec{k}$ space, by the simultaneous presence of altermagnetic order), the aforementioned bands are generically non-degenerate and $d_{\vec{k}}$ have no additional spin quantum number. This is also true (2) in the limit of strong spin-orbit coupling and broken inversion symmetry where the bands are also non-degenerate and $d_{\mathbf{k}}$ carry no additional index. Finally, (3) if the bands are spin degenerate, one should replace $d_{\mathbf{k}} \rightarrow d_{\mathbf{k},\sigma}$ with an additional spin quantum number $\sigma$. Our analysis still applies to this case as well, but we assume for notational simplicity here that the impurities do not carry spin such that $W_{\vec{k},\vec{k}'}$ is not a matrix in spin space.

In this description, the superconducting order parameter is a $\vec{k}$-dependent complex function $\Delta_{\vec{k}}$ coupling to the electrons as
\begin{equation}
    \sum_{\mathbf{k}} d^\dagger_{\mathbf{k}+\frac{\mathbf{q}}{2}} \Delta_{\mathbf{k}} d^{\dagger}_{-\mathbf{k}+\frac{\mathbf{q}}{2}} \,\text{ and }\, \sum_{\mathbf{k}} d^\dagger_{\mathbf{k}+\frac{\mathbf{q}}{2},\uparrow} \Delta_{\mathbf{k}} d^{\dagger}_{-\mathbf{k}+\frac{\mathbf{q}}{2},\downarrow} \label{SCOrderParameter}
\end{equation}
for cases (1,2) and (3), respectively. For later reference, we here also allowed for finite momentum, $\vec{q}\neq 0$, Cooper pairs. Note that $\Delta_{\mathbf{k}} = -\Delta_{-\mathbf{k}}$ for the first, spinless case in \equref{SCOrderParameter} due to Fermi statistics.

As already mentioned above, the gauge transformation~(\ref{gaugetrafro}) of the Bloch states affects $\Delta_{\vec{k}}$. In the presence of TRS in the normal state, $\Theta h_{\vec{k}} \Theta^\dagger = h_{-\vec{k}}$, however, we have $\ket{\phi_{-\vec{k}}} = e^{i\rho_{\vec{k}}}\Theta \ket{\phi_{\vec{k}}}$ which we can use to fix the gauge. Furthermore, TRS favors pairing of Kramers' partners and thus $\vec{q}=0$. For cases (1) and (3), we effectively have $\Theta^2 = \mathbbm{1}$ and can thus choose $e^{i\rho_{\vec{k}}} = 1$ [see SI for the discussion of case (2)]. In this gauge, it holds 
\begin{equation}
    w = t\,\Theta w \Theta^\dagger \, \Rightarrow \, W_{\mathbf{k},\mathbf{k}'} = t(W_{-\mathbf{k},-\mathbf{k}'})^*, \quad t=\pm, \label{TRSOnMatrixElements}
\end{equation}
and it is natural to split a generic disorder matrix element into a non-magnetic ($t=+$) and magnetic ($t=-$) component, 
$W=W^+ + W^-$ with $W_{\mathbf{k},\mathbf{k}'}^{t} = t\left( W_{-\mathbf{k},-\mathbf{k}'}^{t} \right)^*$.
The celebrated Anderson theorem \cite{AndersonPaper,abrikosov1959theory,abrikosov1959superconducting} states that non-magnetic disorder matrix elements, $W=W^+$, do not affect the critical temperature of a BCS-like superconductor with $\Delta_{\vec{k}} = \text{const.}$, $\vec{q}=0$, and assuming $\xi_{\vec{k}} = \xi_{-\vec{k}}$ as also follows from TRS. 

In the absence of TRS in the normal-state Hamiltonian, which is our main interest here, \equref{TRSOnMatrixElements} is in general not valid anymore, such that already intrinsically nonmagnetic impurities ($w = \Theta w \Theta^\dagger$) lead to an effectively magnetic part $W^- \neq 0$, indicating that even a BCS-like state is suppressed by impurities. What is more, it is not even clear how to fix the gauge anymore, giving $\Delta_{\vec{k}} = \text{const.}$ a well-defined meaning. On top of this, the Anderson theorem heavily relies on the parent band dispersion being time-reversal symmetric and we will also analyze how $\xi_{\vec{k}} \neq \xi_{-\vec{k}}$ affects the disorder sensitivity and interferes with the above-mentioned wave function effects.

\begin{figure}[b]
    \centering
    \includegraphics[width=1.0\linewidth]{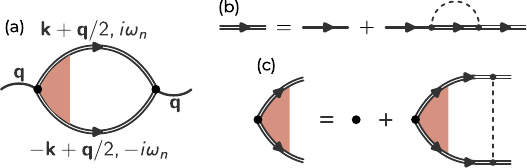}
    \caption{\textbf{Diagrammatics.} (a) shows an exact diagrammatic representation of the disordered particle-particle bubble, expressed in terms of the full Green's function (double line) and the renormalized vertex (colored triangle). In the limit $k_{\mathrm{}{F}} l \gg 1$, the dressed Green's function and the vertex reduce to the noncrossing diagrams shown in (b) and (c), respectively.}
    \label{fig:diagrams}
\end{figure}

To study this, we will assume that the superconducting phase transition is of second order and center our analysis around the disordered particle-particle bubble, which enters the quadratic part of the Ginzburg-Landau expansion.
It is represented diagrammatically in \figref{fig:diagrams}a. If the disorder is sufficiently weak so that the mean free path $l = v_{F} \tau$ satisfies $k_{F}l \gg 1$, single-particle interference effects become negligible. In this regime, diagrams containing crossed impurity lines can be ignored, as they are suppressed by a factor of $(k_{F}l)^{-1}$; the dressed Green's function and vertex correction are then given by the rainbow diagrams and the Cooperon ladder in Fig.~\ref{fig:diagrams}b and c, respectively.

We now briefly discuss the main approximations we will use to obtain analytical results. We first split the energy into the symmetric and antisymmetric parts, $\xi_{\mathbf{k}} =  \tilde{\xi}_{\mathbf{k}} + a_{\mathbf{k}}, \tilde{\xi}_{\mathbf{k}} = ( \xi_{\mathbf{k}} + \xi_{-\mathbf{k}} )/2,  a_{\mathbf{k}} = ( \xi_{\mathbf{k}} - \xi_{-\mathbf{k}} )/2$, and rewrite all summations over $\mathbf{k}$ using the following generalized constant-DOS approximation,
\begin{align}
    \sum_{\mathbf{k}} \ldots \to \int d\varphi_{\mathbf{k}} \tilde{\rho}(\varphi_{\mathbf{k}}) \int d\tilde{\xi} \ldots,
\end{align}
where $\tilde{\rho}(\varphi_{\mathbf{k}})$ is the angular-resolved density of states of the symmetrized energies $\tilde{\xi}_{\mathbf{k}}$ at the associated symmetrized Fermi surface $\tilde{\xi}_{\mathbf{k}} = 0$, the integral over $\tilde{\xi}$ corresponds to summation over the momentum perpendicular to the above-mentioned Fermi surface. We will also assume that $a_{\mathbf{k}}$ and the disorder matrix elements $W_{\mathbf{k},\mathbf{k}'}$ depend only on the angle $\varphi_{\mathbf{k}/\mathbf{k}'}$ and not on $\tilde{\xi}$. 

We begin our analysis by expanding the particle-particle bubble to leading order of $\gamma^2$ to derive a general expression for the dimensionless disorder sensitivity $\zeta$ which captures the deviation of the critical temperature $T_c(\tau)$ from its clean value $T_{c,0}$ according to
\begin{equation}
    \frac{T_{c}(\tau) - T_{c,0}}{T_{c,0}} = - \frac{\pi}{4 T_{c,0} \tau}\zeta + \mathcal{O}(\tau^{-2}), \label{DimensionlessSuppressionOfTc}
\end{equation}
i.e., for large average life times $\tau$ (small $\gamma^2 \propto \tau^{-1}$). The quantity is normalized such that $\zeta = 1$ for the usual Abrikosov-Gorkov law (magnetic impurities in BCS superconductor) and $\zeta = 1/2$ for scalar impurities in an unconventional superconductor \cite{PhysRevResearch.2.023140}.

Allowing for an arbitrary order parameter $\Delta_{\mathbf{k}}$, general complex $W_{\vec{k},\vec{k}'}$, and $a_{\vec{k}} \neq 0$, we obtain the compact result
\begin{equation}
    \zeta = \frac{\sum_{t=\pm} \sum_{\mathbf{k},\mathbf{k}'}^{\mathrm{FS}} \left( \bigl| C_{\mathbf{k},\mathbf{k}',f_1}^{t} \bigr|^2 - t  \bigl| C_{\mathbf{k},\mathbf{k}',f_2}^{t} \bigr|^2 \right)}{ 4 \sum_{\mathbf{k},\mathbf{k}'}^{\mathrm{FS}}  \left| W_{\mathbf{k},\mathbf{k}'} \right|^2 \sum_{\mathbf{k}}^{\mathrm{FS}} {|\Delta_{\mathbf{k}}|}^2 f_3(\mathbf{k}) }, \label{eq:general_zeta}
\end{equation}
where $\sum_{\mathbf{k}}^{\mathrm{FS}} \ldots = \tilde{\nu}^{-1} \int d\varphi_{\mathbf{k}}\tilde{\rho}(\varphi_{\mathbf{k}})\ldots$, $\tilde{\nu} = \int d\varphi_{\mathbf{k}}\tilde{\rho}(\varphi_{\mathbf{k}})$, is a normalized Fermi-surface average and
\begin{subequations}\begin{align}
    C^{\pm}_{\mathbf{k},\mathbf{k}',f_1} &= W_{\mathbf{k},\mathbf{k}'} \Delta_{\mathbf{k}'} f_{1}(\mathbf{k}') \pm f_{1}(\mathbf{k}) \Delta_{\mathbf{k}} W_{-\mathbf{k},-\mathbf{k}'}^*,\\
    C^{\pm}_{\mathbf{k},\mathbf{k}',f_2} &= f_{2}(\mathbf{k},\mathbf{k}') \left[ W_{\mathbf{k},\mathbf{k}'} \Delta_{\mathbf{k}'}  \pm \Delta_{\mathbf{k}} W_{-\mathbf{k},-\mathbf{k}'}^{*} \right].
\end{align}\end{subequations}
Writing $W=W^+ + W^-$ as before, one can think of $C^{\pm}_{f_{1,2}}$ as the sum of the commutator of $\Delta$ with $W^\mp$ and of the anti-commutator of $\Delta$ with $W^\pm$, with $\Delta$ and $W$ dressed by
\begin{subequations}\begin{align}
    f_1(\mathbf{k})&=\frac{1}{\sqrt{2}\cosh(a_{\mathbf{k}}/T_{c,0})},\\
    f_2(\mathbf{k}, \mathbf{k}') &= \left[  \frac{a_{\vec{k}} \tanh[\frac{a_{\mathbf{k}}}{2T_{c,0}}] - a_{\vec{k}'} \tanh[\frac{a_{\mathbf{k}'}}{2T_{c,0}}]}{(a_{\vec{k}}^2 - a_{\vec{k}'}^2)/T_{c,0}} \right]^{1/2},
\end{align}\end{subequations}
respectively. 
These factors are the spectral weights entering the perturbative expansion of the particle-particle bubble---$f_{1}$ comes from the self-energy correction and $f_2$ from the diagram with one disorder rung. Finally, the function $f_{3}$ in \equref{eq:general_zeta} is proportional to the temperature derivative of the bare particle-particle bubble; for small $a_{\vec{k}}$, it behaves as $f_3(\vec{k}) \sim 1 + \mathcal{O}(a_{\vec{k}}^2)$ and its exact form can be found in the SI. 

Note that \equref{eq:general_zeta} readily reproduces the Anderson theorem: for $a_{\vec{k}}\rightarrow 0$, we see that $f_1=f_2=1/\sqrt{2}$ such that, for non-magnetic impurities, $\zeta$ is proportional to the commutator of $\Delta$ and $W$, which vanishes for constant $\Delta_{\vec{k}}$. Beyond this simple limit, \equref{eq:general_zeta} encodes wave function effects and the consequences of $\xi_{\vec{k}} \neq \xi_{-\vec{k}}$ as well as their complex interplay. Equation~(\ref{eq:general_zeta}) already indicates, as we will also explicitly see below, that even $\zeta < 0$  is possible, i.e., an enhancement of $T_c$ with disorder. In the remainder of the paper, we will scrutinize these effects, different limits, and applications, both based on \equref{eq:general_zeta} and going beyond leading order in $\gamma^2$.

\subsection{Quantum geometric pair breaking}
We begin by isolating the effects of the wave functions and set $a_{\mathbf{k}} \rightarrow 0$. From the general expression for the suppression factor in \equref{eq:general_zeta}, we arrive at
\begin{align}
    \zeta &= \frac{\sum_{\mathbf{k},\mathbf{k}'}^{\mathrm{FS}} \left| W_{\mathbf{k},\mathbf{k}'} \Delta_{\mathbf{k}'} - \Delta_{\mathbf{k}} W^*_{-\mathbf{k},-\mathbf{k}'} \right|^2}{4 \sum_{\mathbf{k},\mathbf{k}'}^{\mathrm{FS}} \left|W_{\mathbf{k},\mathbf{k}'}\right|^2 \sum_{\mathbf{k}}^{\mathrm{FS}} \left|\Delta_{\mathbf{k}}\right|^2 }, \label{ZetaSimplified}
\end{align}
which is consistent with the expression derived in \cite{PhysRevResearch.2.023140} and agrees with \cite{doi:10.1126/sciadv.aay6502} for $\Delta_{\vec{k}}=\text{const.}$, in the range of applicability of the results of \cite{doi:10.1126/sciadv.aay6502}. 
Before applying \equref{ZetaSimplified} to some concrete examples, we will discuss how it can be used to define a natural gauge for a disordered superconductor without TRS. We formulate it as the gauge which, for \textit{fixed} $\Delta_{\vec{k}}$ and $w$, yields minimal $\zeta$. Since the denominator of \equref{ZetaSimplified} is gauge invariant, we can focus on the numerator,
\begin{align}
    I = \sum_{\mathbf{k},\mathbf{k}'}^{\mathrm{FS}} \left| W_{\mathbf{k},\mathbf{k}'} \Delta_{\mathbf{k}'} - \Delta_{\mathbf{k}} W^*_{-\mathbf{k},-\mathbf{k}'} \right|^2  \geq 0. \label{IExpression}
\end{align}
This functional can be further split into two contributions, $I = 2 \sum_{\mathbf{k},\mathbf{k}'}^{\mathrm{FS}}  |\Delta_{\mathbf{k}}|^2 |W_{\mathbf{k},\mathbf{k}'}|^2 - 2 \tilde{I}$; the first term is gauge-invariant, while the second part acquires momentum-dependent phase factors under \equref{gaugetrafro}. This allows us to write it explicitly as a function of $\bar{\alpha}_{\mathbf{k}} = \alpha_{\mathbf{k}} + \alpha_{-\mathbf{k}}$ as
\begin{equation}
        \tilde{I} \to \tilde{I}[\bar{\alpha}_{\vec{k}}] = \sum_{\mathbf{k},\mathbf{k}'}^{\mathrm{FS}} \Delta_{\mathbf{k}}^* W_{\mathbf{k},\mathbf{k}'} W_{-\mathbf{k},-\mathbf{k}'} \Delta_{\vec{k}'} e^{i(\bar{\alpha}_{\mathbf{k}'} - \bar{\alpha}_{\mathbf{k}})}. \label{ItildeOptimization}
\end{equation}
To simplify the presentation, we will first study the spin degenerate case (3), before discussing (1,2) below. While one can in principle use any $\Delta_{\vec{k}}$, the most natural approach for case (3) is to define the optimal gauge as the one that maximizes $\tilde{I}[\bar{\alpha}_{\vec{k}}]$ for $\Delta_{\vec{k}} = \Delta_0$, since it most directly connects to the case with TRS. In fact, in the presence of TRS in the normal state, there are phase factors $e^{i\rho_{\vec{k}}} = e^{i \rho_{-\vec{k}}}$ such that $\ket{\phi_{-\vec{k}}} = e^{i\rho_{\vec{k}}}\Theta \ket{\phi_{\vec{k}}}$ and it is straightforward to see that \equref{ItildeOptimization} is maximized by $e^{i\bar{\alpha}_{\mathbf{k}}}=e^{-i\rho_{\vec{k}}}$ (thus, effectively, $e^{i\rho_{\vec{k}}} \rightarrow 1$) for any non-magnetic impurity with maximum value given by $\tilde{I}[\bar{\alpha}_{\mathbf{k}}^{\mathrm{opt}}] = |\Delta_0|^2 \sum_{\mathbf{k},\mathbf{k}'}^{\mathrm{FS}} |W_{\vec{k},\vec{k}'}|^2$. As such, our optimization procedure recovers the conventional phase choice for superconductors with TRS discussed above, with $I$ further reaching its lower bound, $I_{\text{opt}} = 0$, in line with Anderson's theorem ($\zeta_{\text{opt}}=0$). 

Without TRS, however, the optimization problem is in general ``frustrated'' in the sense that $I_{\text{opt}} > 0$ and, thus, $\zeta_{\text{opt}} > 0$. Loosely speaking, this is expected since for $N_k$ values of $\vec{k}$ in \equref{IExpression}, there are $N_k^2$ values of $W_{\mathbf{k},\mathbf{k}'}$ but only $N_k$ phases to optimize. Physically, $\zeta_{\text{opt}} > 0$ means that, irrespective of the phases of the superconducting order parameter, the underlying TRS-breaking part of the quantum geometry induces a pair breaking effect, which we dub ``quantum geometric pair breaking''.

To provide a real-space interpretation of this effect, let us rewrite the pairing part of the Hamiltonian (\ref{SCOrderParameter}) in the original real-space basis,
\begin{align}
    \sum_{\mathbf{k}} \Delta_{\mathbf{k}} d^{\dagger}_{\mathbf{k},\uparrow} d^{\dagger}_{-\mathbf{k},\downarrow} = \sum_{\mathbf{r},\mathbf{x},\alpha,\alpha'} c^\dagger_{\mathbf{r}-\frac{\mathbf{x}}{2},\alpha,\uparrow} D_{\alpha,\alpha'}(\mathbf{x}) c^\dagger_{\mathbf{r}+\frac{\mathbf{x}}{2},\alpha',\downarrow}. \label{RealSpacePairing}
\end{align}
Here and in the following, we absorbed the projection onto the Fermi surface into the redefinition of the order parameter, $\Delta_{\mathbf{k}} \to \Delta_{\mathbf{k}} \delta(\xi) / \nu$, with weight located around the Fermi surface. The real-space form of the order parameter in \equref{RealSpacePairing} is given by
\begin{align}
    D_{\alpha,\alpha'}(\vec{x}) = \sum_{\vec{k}} \Delta_{\vec{k}} \phi_{\vec{k},\alpha} \phi_{-\vec{k},\alpha'} e^{i\mathbf{kx}} \label{eq:Dx_def}
\end{align}
and can be thought of as the analogue of a Wannier state for Cooper pairs, but constructed from the (for $|\Delta_{\vec{k}}| \neq \text{const.}$ unnormalized) two-particle Bloch-like state $(U_{\vec{k}})_{\alpha,\alpha'} = \Delta_{\mathbf{k}} \phi_{\mathbf{k},\alpha} \phi_{-\mathbf{k},\alpha'}$. 
As $D$ is complex and matrix-valued, we probe its spatial behavior using a real, scalar measure of the form
\begin{align}
    C_{v}(\mathbf{x}) := \mathrm{Tr} \left[ D^\dagger(\vec{x}) v^* D(\vec{x}) v \right], \quad v^\dagger = v, \label{eq:Cx_def}
\end{align}
which picks up different components of $D$ depending on the choice of $v$ and becomes the Frobenius norm for $v=\mathbbm{1}$. 
Remarkably, choosing $v=w$, we find $C_{w}(\mathbf{x} = 0) = \tilde{I}$, meaning that the minimization of the superconducting suppression rate $\zeta$ is equivalent to finding the gauge in which the Cooper pair wavefunction is maximally localized in the sense of having largest intra-unit-cell weight $C_{w}(\mathbf{x} = 0)$ or, equivalently, smallest $\sum_{\vec{x}\neq 0}C_{w}(\mathbf{x})$~\footnote{Assuming the latter is finite.}. 
Although the derived localization functional differs from that of the usual Wannierization problem centered around the minimization of the variance of the electron position \cite{RevModPhys.84.1419}, this result formally connects the gauge choice in disordered superconductors and the disorder sensitivity of the pairing state to finding localized Wannier-like states. It also provides another perspective on the role of the quantum geometry for the stability of superconductivity in TRS-broken systems: similar to how quantum geometry constrains the localization of Wannier states in Bloch bands, non-trivial $\vec{k}$ dependencies in the composite state $U_{\vec{k}}$ induce spatial extend in $C_w$ and, thus, $\zeta_{\text{opt}} > 0$. 

\begin{figure*}[tb]
    \centering
    \includegraphics[width=1.0\linewidth]{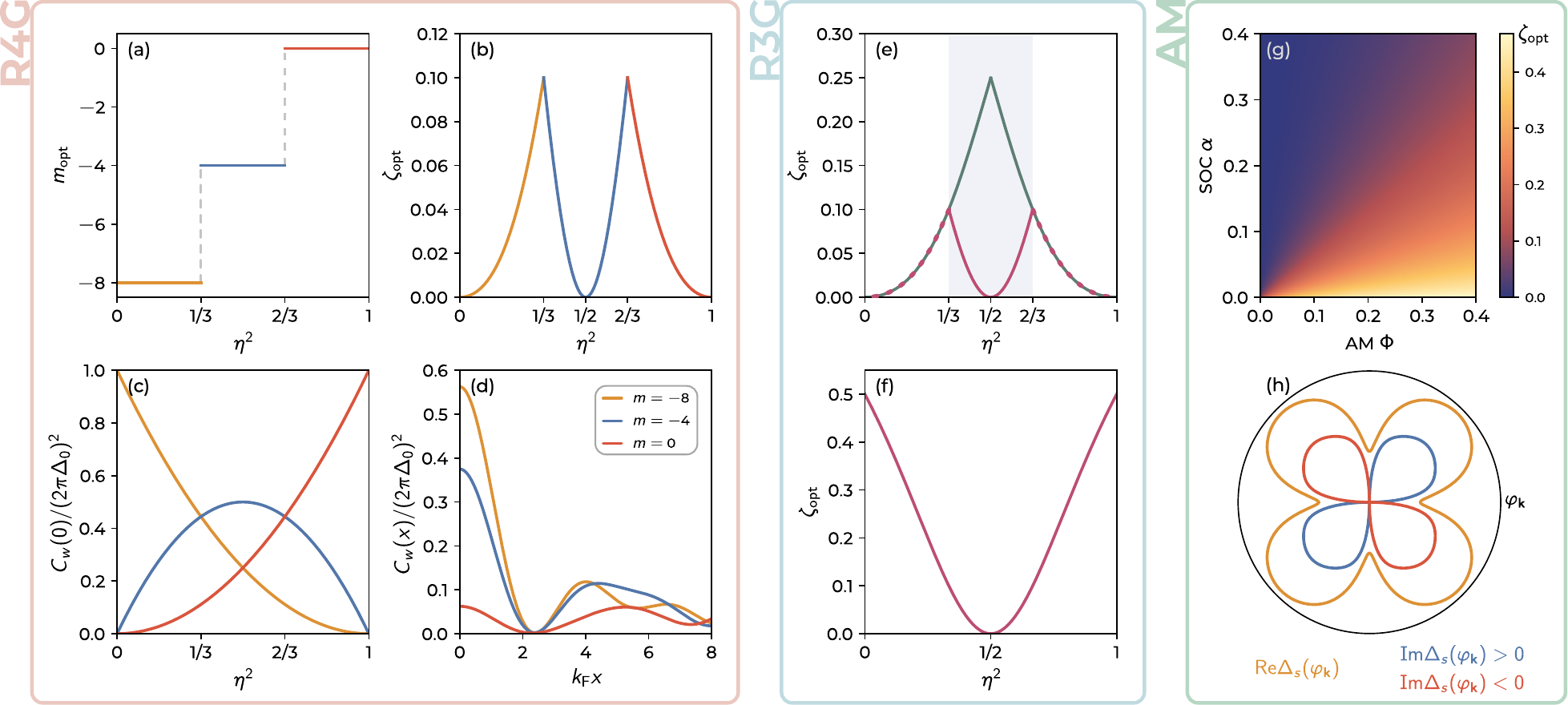}
    \caption{\textbf{Gauge choice, disorder sensitivity, and localization of Cooper-pair wave function.} Here, (a-d), (e-f), and (g-h) refer to $n=4$ (R$4$G) in \equref{MatrixElements}, $n=3$ (R$3$G), and altermagnetism, respectively. 
    The dependence of (a) $m_{\mathrm{opt}}$ defining the optimal gauge and superconducting Chern numbers, and of the corresponding $\zeta_{\mathrm{opt}}$ (b) on $\eta^2$. (c) shows the normalized measure of the localization of the Cooper pair wavefunction $C_{w}(\mathbf{x}=0)$, and (d) the spatial profile of this function for $\eta^2=1/4$ for the indicated $m$. As opposed to $n=4$ in (a-d), we find for $n=3$ (e) that $\zeta_{\text{opt}}$ differs in the shaded region between gauge (green) and $\Delta_{\vec{k}}$ (red) optimization, where the most stable superconductor is a triplet state ($\Delta_{\vec{k}} = -\Delta_{-\vec{k}}$). (f) Shows $\zeta_{\mathrm{opt}}$ for the spinless case in R$3$G, where again optimal gauge choice and superconductor coincide, revealing an emergent Anderson theorem for $\eta^2=1/2$. (g) shows the optimal $\zeta$ for a checkerboard lattice model of altermagnetism as a function of spin-orbit coupling $\alpha$ and altermagnetic strength $\Phi$. To illustrate the symmetry of the pairing state, which is the same for all $\alpha,\Phi \neq 0$, we show the angular dependence of the singlet component $\Delta_s$ of the optimal pairing state for $\alpha=\Phi=0.5$ in (h); for clarity, the imaginary part is multiplied by 15. There is also a non-unitary, in-plane triplet component (not shown).}
    \label{fig:optimization} 
\end{figure*}

Note that our discussion here of quantum geometric pairing can be straightforwardly generalized to pairing at finite center of mass momentum $\vec{q} \neq 0$. To simplify the discussion, we however leave this to the SI and continue here with an illustration of these aspects using the concrete form 
\begin{equation}
    W_{\mathbf{k},\mathbf{k}'} = W_0\left[\eta^2 + (1 - \eta^2) e^{in(\varphi_{\mathbf{k}'} - \varphi_{\mathbf{k}})}\right] \label{MatrixElements}
\end{equation}
of the disorder matrix elements, where $e^{i \varphi_{\mathbf{k}}} = k/|k|$, $k=k_x+ik_y$. As $W_0$ drops out in \equref{ZetaSimplified}, these matrix elements are effectively parameterized by just two non-trivial constants $\eta \in [0, 1]$ and $n\in\mathbb{Z}$. Equation~(\ref{MatrixElements}) arises, for instance, from the minimal, isotropic two-band model of rhombohedral $n$-layer graphene (R$n$G) where only the dominant degrees of freedom---the $A_1$ (upper component of Hamiltonian) and $B_n$ (lower component)---are kept~\cite{Model_Koshino},
\begin{equation}
    h_{\vec{k}} =
    \begin{pmatrix}
        u_0 - \mu & t_0 (k_x - i k_y)^n \\ t_0 (k_x + i k_y)^n & -u_0 - \mu
    \end{pmatrix}, \label{HamiltonianRTG}
\end{equation}
taking the (microscopically non-magnetic) disorder matrix elements to be diagonal $w = \mathrm{diag}(w_A, w_B)$, and using the gauge where the Bloch states $\ket{\phi_{\vec{k}}}$ of the active bands transform trivially under rotations around $A_1$, $C_{3z}^{A_1} : d_\mathbf{k}^\dagger \to d^\dagger_{C_{3z} \mathbf{k}}$ \cite{TheSTMPaper}. Note the value of $\eta$ is tunable by the displacement field, $u_0$ in \equref{HamiltonianRTG}, and depends on $w_A/w_B$.

It is straightforward to show that the solutions of the stationary equation $\delta_{\bar{\alpha}_{\vec{k}}}\tilde{I}[\bar{\alpha}_{\vec{k}}] = 0$ are of the form $\bar{\alpha}_{\mathbf{k}} = m \varphi_{\mathbf{k}}, m \in \mathbbm{Z}$. In \figref{fig:optimization}a-b, we present the $\eta$ dependence of the optimal $m$ that corresponds to the global maximum of $\tilde{I}[\bar{\alpha}_{\vec{k}}]$ and of the corresponding suppression factor $\zeta_{\mathrm{opt}}$, taking $n=4$ as an example. We see that, as anticipated, the matrix elements (via $\eta$) crucially determine the optimal gauge and that---away from the fine-tuned points $\eta^2=0,1/2,1$---$\zeta_{\mathrm{opt}} > 0$, revealing non-trivial quantum geometric pair breaking. For the simple matrix elements in \equref{MatrixElements} and a circular Fermi surface, the associated Cooper-pair wave function $D$ and, thus, $C$ are straightforwardly computed analytically (see SI for expression). As can be seen in \figref{fig:optimization}c-d, where we illustrate them for the three solutions $m=0,-4,-8$ that dominate for some $\eta$, we find that the optimal gauge indeed corresponds to the one with maximal $C_w(0)$.

Finally, we mention that there is a physically rather different but mathematically very closely related perspective: instead of using $\zeta$ to define a gauge, we can ask the question which superconductor, i.e., which $\Delta_{\vec{k}}$ in some given gauge, will be most stable. This is equivalent to optimizing $I$ in \equref{IExpression} with respect to $\Delta_{\vec{k}}$ at fixed $\sum_{\mathbf{k}}^{\mathrm{FS}} |\Delta_{\mathbf{k}}|^2$ and, thus, to finding the smallest eigenvalue of the positive semi-definite operator $\mathcal{M}$ defined via $I = \sum_{\mathbf{k},\mathbf{k}'}^{\mathrm{FS}} \Delta^*_{\vec{k}} \mathcal{M}_{\vec{k},\vec{k}'} \Delta^{\phantom{*}}_{\vec{k}'}$. 

We note that, while this optimization procedure now also involves changing $|\Delta_{\vec{k}}|$ and not only the phases, it is in general also frustrated leading to $\zeta_{\text{opt}} > 0$, i.e., quantum geometry leads to pair breaking for \textit{any} superconductor. An example is again provided by the matrix elements in \equref{MatrixElements} with $n=4$. Here, it turns out that the optimal solutions obey $|\Delta_{\vec{k}}| = \text{const.}$ and $\Delta_{\vec{k}}=\Delta_{-\vec{k}}$, such that we are effectively back to optimizing the phases only and the results of \figref{fig:optimization}a-d still apply. The only difference is that one should think of $m_{\mathrm{opt}}$ as defining the superconducting order parameter via $\Delta_{\vec{k}} = e^{i\varphi_{\vec{k}} m_{\mathrm{opt}}}$. Interpreting \equref{MatrixElements} as coming from R4G as explained above, we note that in the gauge used, $m_{\mathrm{opt}}$ is the Chern number of the superconductor \cite{TheSTMPaper,ChristosEnergetics}. Therefore, \figref{fig:optimization}a shows that, without TRS, quantum geometric effects in the disorder matrix elements can make topologically non-trivial superconductors more protected against microscopically non-magnetic disorder than their trivial counterparts. 

Naturally, finding the optimal gauge is not always the same as optimizing $\Delta_{\vec{k}}$. For instance, in \figref{fig:optimization}e, we find $\Delta_{\vec{k}} = -\Delta_{-\vec{k}}$ as the most stable superconductor for $1/3 < \eta^2 < 2/3$ in the case of $n=3$ in \equref{MatrixElements}, i.e., R$3$G. Therefore, $\zeta_{\text{opt}}$ from the gauge optimization is larger than that from the superconducting order parameter optimization. This shows that quantum geometric pair breaking can lead to the surprising result that there is a triplet superconductor ($\Delta_{\vec{k}} = -\Delta_{-\vec{k}}$) that is more stable than any spin singlet superconductor ($\Delta_{\vec{k}} = \Delta_{-\vec{k}}$).

\subsection{Non-degenerate Fermi surfaces}
For the cases (1,2) with non-degenerate Fermi surfaces, the superconducting order parameter in \equref{SCOrderParameter} obeys $\Delta_{\vec{k}} = -\Delta_{-\vec{k}}$. As this is inconsistent with $\Delta_{\vec{k}} = \Delta_0$, one has to choose another $\Delta_{\vec{k}}$ to fix the gauge. While other possibilities are conceivable, we propose $\Delta_{\vec{k}} = \Delta_0 e^{i\varphi_{\vec{k}}}$ as a natural option due to its connection to the limit with TRS. Using $\ket{\phi_{-\vec{k}}} = e^{i\rho_{\vec{k}}}\Theta \ket{\phi_{\vec{k}}}$, one finds that $\tilde{I}[\bar{\alpha}_{\vec{k}}]$ in \equref{ItildeOptimization} is maximized for $e^{i\bar{\alpha}_{\mathbf{k}}}=e^{-i(\rho_{\vec{k}}+\varphi_{\vec{k}})}$ yielding $I_{\text{opt}} = 0$ (Anderson theorem); note that this solution is only possible for case (2), where $e^{i\rho_{\vec{k}}}$ is odd in $\vec{k}$. For case (1), the gauge choice optimization problem is already frustrated with TRS in the normal state and there is generically no Anderson theorem ($I_{\text{opt}} > 0$), for any gauge choice or superconductor.

This, however, does \textit{not} mean that the superconductor always effectively behaves like an unconventional state with $\zeta=1/2$. To illustrate this, let us again consider the matrix elements in \equref{MatrixElements} for $n=3$, i.e., R$3$G. For $\Delta_{\vec{k}} = \Delta_0 e^{i\varphi_{\vec{k}}}$, we now find $\bar{\alpha}_{\mathbf{k}} = -4 \varphi_{\mathbf{k}}$ for all $\eta$ with $\zeta_{\text{opt}}$ shown in \figref{fig:optimization}f. We can see that $\zeta_{\text{opt}}$ only reaches $1/2$ for $\eta=0,1$ and is smaller in between, with an emergent Anderson theorem for $\eta=1/2$. 
Similar to our discussion above, we can also reinterpret the gauge optimization as a search for the most stable superconductor, which in the current case turns out to be $\Delta_{\vec{k}} = \Delta_0 e^{-3i\varphi_{\vec{k}}}$, for all $\eta$.  
We emphasize that $\zeta_{\text{opt}}<1/2$ for $\eta\neq 0,1$ is a consequence of quantum geometry since $W_{\vec{k},\vec{k}'} = \text{const.}$ (as realized for $\eta=1$) would always imply $\zeta_{\text{opt}}=1/2$. 

As anticipated above, the physics and concepts discussed here also find a very natural application in altermagnets. To illustrate this with a concrete example, we use a two-sublattice model on the checkerboard lattice (see SI) and study it as a function of the altermagnetic order parameter $\Phi$ and spin-orbit coupling strength $\alpha$. We focus on the parameter regime with two active Fermi surfaces, which are split for generic parameters; despite the broken TRS, the respective band energies are still even, $\xi_{\vec{k}} = \xi_{-\vec{k}}$, which follows from a two-fold rotational symmetry. In line with our discussion thus far, we further focus on pairing of the states on the Fermi surfaces at $\vec{k}$ and $-\vec{k}$. This is natural in the weak-coupling limit, where only these types of states exhibit a Cooper logarithm and should thus be favored, as long as the associated effective Cooper-channel interaction is attractive.

In \figref{fig:optimization}g, we show $\zeta$ for the optimal $\Delta_{\vec{k}}$ using simple non-magnetic impurities with $w=\mathbbm{1}_{4\times 4}$, where quantum geometry plays a dual role: on the one hand, in the limit of small $\Phi/\alpha$, one can think of $\Phi$ as breaking the spinfull TRS ($\Theta^2=-\mathbbm{1}$) such that the normal-state wave functions $\ket{\phi_{\vec{k}}}$ dress the impurities to also develop an effectively magnetic component, leading to $\zeta_{\text{opt}} > 0$. On the other hand, for $\alpha/\Phi \ll 1$, the system is close to an ``orbital'' TRS, $\Theta^2 =\mathbbm{1}$. The fact that $\zeta_{\text{opt}}$ stays below $1/2$ for small $\alpha$ (and finite $\Phi$) is---just as in the case of spin-polarized R$3$G discussed above---a result of non-constant $|W_{\vec{k},\vec{k}'}|$. The pairing state this most stable transforms trivially under all (magnetic) point symmetries, see \figref{fig:optimization}h.
While the model and state are different, we note that our non-zero value of $\zeta$ for small $\alpha/\Phi$ is in line with the suppression of superconductivity in \refcite{PhysRevB.108.054510} at larger altermagnetic strength. We further emphasize that the spinful TRS in a collinear antiferromagnet will ensure that $\zeta_{\text{opt}}=0$, in contrast to the altermagnet.

\subsection{Concrete pairing states}
Finally, one can also use $\zeta$ in \equref{ZetaSimplified} to study the stability of any fixed candidate superconductor as a function of system parameters. We will again turn to R4G as an example (with or without spin polarization), with Hamiltonian in \equref{HamiltonianRTG}, microscopically non-magnetic  $w = \mathrm{diag}(w_A, w_B)$, and the $C_{3z}^{A_1}$-symmetric gauge \cite{TheSTMPaper}. We first consider $\Delta_{\mathbf{k}} = \Delta_0$, which transforms under the trivial irreducible representation of $C_{3z}^{A_1}$ and corresponds to the ``achiral'' \cite{ChristosEnergetics} superconductor with zero Chern number. Equation~(\ref{ZetaSimplified}) readily yields a suppression rate of $\zeta=\alpha^2/(2(1 + \alpha^2))$, where $\alpha = (w_B/ w_A) (\gamma_0/(\sqrt{u_0^2+\gamma_0^2}+u_0))^2$, $\gamma_0 = t_0 k_F^4$.
Let us consider two limiting cases. First, when all impurities are located on the $A_1$-sublattice, $\alpha=0$ and superconductivity is completely unaffected by the presence of impurities. Relative to this superconducting order parameter, disorder is nonmagnetic as $W_{\vec{k},\vec{k}'}$ in \equref{MatrixElements} become real (for $\eta\rightarrow 1$), and we therefore recover the Anderson theorem result. When $w_B / w_A \to \infty$, however, we get $\zeta = 1/2$ which is a signature of unconventional pairing. Intuitively, this follows from the phase modulation of the intra-valley Cooper pair wavefunction within the unit cell, which is the reason why the order parameter transforms under a non-trivial irreducible representation of $C_{3z}^{B_4}$ \cite{TheSTMPaper} and thus effectively acts as unconventional pairing from the perspective of impurities on the $B_4$ sublattice. 

For any ``chiral'' ($m>0$) or ``anti-chiral'' ($m<0$) pairing state $\Delta_{\mathbf{k}} = \Delta_0 e^{\pm i m \varphi_{\mathbf{k}}}$ with $0<|m|<4$, the suppression rate is $\zeta = 1/2$, independent of the ratio $w_A / w_B$. Intuitively, this is because the order parameter behaves as an unconventional superconductor on both $A_1$ and $B_4$ sublattices. Note this is consistent with \figref{fig:optimization}a-b where $\zeta_{\mathrm{opt}} < 1/2$ and $m_{\text{opt}}=0$ or $|m_{\text{opt}}| \geq 4$.

\subsection{Energy effect}
Having discussed the wave-function effects in detail while assuming $\xi_{\vec{k}} = \xi_{-\vec{k}}$, we next focus on the consequences of the TRS-breaking part of the dispersion, $a_{\mathbf{k}} = (\xi_{\mathbf{k}} - \xi_{-\mathbf{k}})/2$, for the disorder sensitivity. To keep the discussion focused, let us assume that $\Delta_{\vec{k}} = \Delta_0$ for $\vec{k}$ around the Fermi surface by properly adjusting the gauge.  

We find it instructive to first identify the effect of $a_{\mathbf{k}}$ on the clean superconductor. At zero temperature, the particle-particle bubble is proportional to $\sum_{\mathbf{k}}^{\mathrm{FS}} \log( 1 + \omega_D^2/a^2_{\mathbf{k}})$, where $\omega_D$ is the cutoff frequency introduced to regularize the usual logarithmic divergence at the upper limit.
For fixed $\omega_D$, this integral is finite as long as $a_{\vec{k}}$ vanishes at most at isolated points on the Fermi surface; therefore, the emergence of superconductivity requires finite interaction strength, making the SC a strong-coupling instability, and $a_{\mathbf{k}}$ can be thought of as ``kinetic pair-breaking''.

Moving to the disordered SC, we first note that the disorder sensitivity factor $\zeta$ in \equref{ZetaSimplified} is always non-negative when $a_{\mathbf{k}} = 0$, and in the chosen gauge determined only by the magnetic part of the disorder $W_{\mathbf{k},\mathbf{k}'}^{-}$ which is a hallmark of Anderson's theorem. However, the general form in \equref{eq:general_zeta} can lead to either sign. To make this apparent, and since wave-function-related pair breaking was already discussed above, let us assume fully non-magnetic ($W=W^+$) disorder. Expanding \equref{ZetaSimplified} to leading order in $a_{\mathbf{k}}$, we get
\begin{align}
    \zeta = - \frac{ \sum_{\mathbf{k},\mathbf{k}'}^{\mathrm{FS}} \left| W_{\mathbf{k},\mathbf{k}'} \right|^2 (a_{\vec{k}}^2 + a_{\vec{k}'}^2)}{48 T_{c,0}^2 \sum_{\mathbf{k},\mathbf{k}'}^{\mathrm{FS}}  \left| W_{\mathbf{k},\mathbf{k}'} \right|^2} \leq 0,
\end{align}
i.e., superconductivity is enhanced by disorder. Intuitively, this enhancement can be understood as follows: while preserving the coherence between electrons with $\mathbf{k}$ and $-\mathbf{k}$, scattering off non-magnetic disorder (similar to scattering by phonons \cite{6wxh-p4mc}) also averages $a_{\mathbf{k}}$ and suppresses its pair-breaking effect. 
A related, but not identical enhancement with disorder was recently discussed in a model of altermagnetism \cite{PhysRevB.111.L100502}: as opposed to our discussion here, they studied a superconductor that is incompatible with the altermagnetic order parameter and, thus, suppressed by it (and enhanced by averaging it to zero). Note that our enhancement is also distinct from that discussed in \refcite{PhysRevLett.80.5413}.

Our intuitive picture is further consistent with the behavior in the opposite limit of strong disorder (while still keeping $k_{F}l \gg 1$), where we find
\begin{align}
    \frac{\tilde{T}_{c,0} - T_c(\tau)}{\tilde{T}_{c,0}} = \frac{\pi \tau}{2 \tilde{T}_{c,0}} \sum_{\mathbf{k}}^{\mathrm{FS}} a_{\mathbf{k}}^2 \frac{\sum_{\mathbf{k}_1,\mathbf{k}_1'}^{\mathrm{FS}} \left|W_{\mathbf{k}_1,\mathbf{k}_1'}\right|^2}{\sum_{\mathbf{k}'}^{\mathrm{FS}}\left|W_{\mathbf{k},\mathbf{k}'}\right|^2}  + \mathcal{O}(\tau^2) \label{eq:critical_temperature_enhancement}.
\end{align}
Here, $\tilde{T}_{c,0}$ is the critical temperature of the clean superconductor when replacing $\xi_{\vec{k}}$ with the symmetrized dispersion $\tilde{\xi}_{\mathbf{k}}$. Thus, the large-disorder limit is independent of the exact behavior of the antisymmetric part of the energies and is determined purely by $\tilde{\xi}_{\vec{k}}$. Since $T_{c,0} < \tilde{T}_{c,0}$, non-magnetic disorder enhances the critical temperature also for $\tau \rightarrow 0$ (as long as localization corrections can be neglected). 
    
\begin{figure}[tb]
    \centering
    \includegraphics[width=1.0\linewidth]{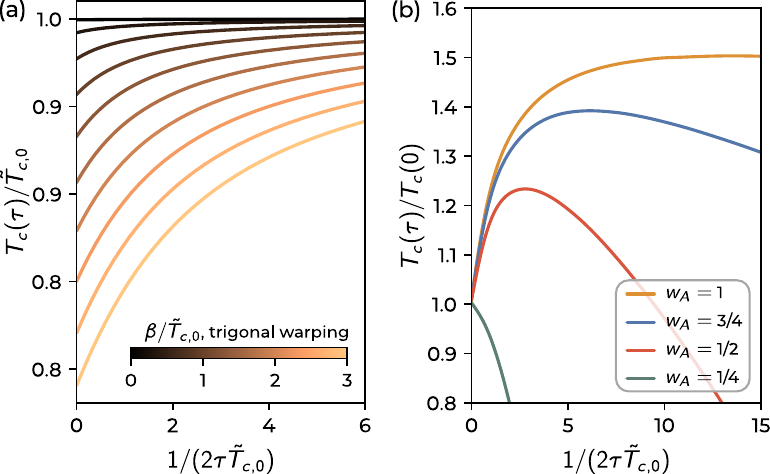}
    \caption{\textbf{Kinetic effects and their interplay with quantum geometry.} (a) Dependence of the critical temperature on the inverse lifetime for different values of the trigonal warping amplitude in the simplified RnG dispersion, which the kinetic pair-breaking $a_{\mathbf{k}} = (\xi_{\mathbf{k}} - \xi_{-\mathbf{k}})/2$. (b) Illustrates the interplay of kinetic and geometric effects on the behavior of the critical temperature in the full continuum R4G model for different disorder impurities $w_A^2 + w_B^2 = 1$. }
    \label{fig:kinetics_and_interplay}
\end{figure}

To illustrate the discussed effect with a concrete example, we consider a simplified dispersion, $\xi_{\mathbf{k}} = |\mathbf{k}|^2/(2m) + \beta \cos(3\varphi) - \mu$, of R$n$G, where the trigonal warping parameter $\beta$ captures the broken TRS in the kinetics. In \figref{fig:kinetics_and_interplay}a, we present the dependence of the critical temperature on the inverse lifetime, for different values of the trigonal warping amplitude $\beta$. First, we note that the critical temperature is suppressed with increasing trigonal warping $\beta$, illustrating the fact that the antisymmetric part of the energy is indeed pair-breaking. For all values of $\beta$, we observe an enhancement of the critical temperature and all curves approach the same limit as $\tau^{-1} \to \infty$ in agreement with \equref{eq:critical_temperature_enhancement}.

\subsection{Composite effect}
We finally consider the interplay between the two discussed effects---quantum geometric and kinetic pair breaking. For concreteness, we apply it to the full 8-band continuum model~\cite{Model_Koshino} of R$4$G (using the parameters of \cite{rtg_chiral_sc_experiment}) and focus on $\Delta_{\vec{k}} = \Delta_0$ in the above-mentioned gauge where $C_{3z}^{A_1}$ acts trivially. Furthermore, we again assume that the impurities are located only on the dominant $A_1$ and $B_4$ sublattice and are microscopically non-magnetic, $w = \mathrm{diag}(w_A, w_B), w_Aw_B \geq 0$, normalized such that $w_A^2 + w_B^2 = 1$. 

In \figref{fig:kinetics_and_interplay}b, we show the dependence of the critical temperature on the disorder strength for different disorder configurations. Starting with the case $w_B = 0$, we recover the enhancement of superconductivity. This naturally follows from $W$ being real [in analogy to to $\eta = 1$ in \equref{MatrixElements}] and we are back to the discussion of the previous subsection.
When both $w_A$ and $w_B$ are nonzero, the critical temperature exhibits a non-monotonic dependence on the disorder strength. At small $\tau^{-1}$, the non-magnetic component enhances superconductivity, however, once the critical temperature reaches its maximum, the effective magnetic contribution of the disorder takes over and begins to suppress it, eventually driving it to zero. For sufficiently small $w_A$, however, the role of the magnetic contribution in $W$ starts to dominate \equref{eq:general_zeta} such that $\zeta > 0$ and the critical temperature decreases monotonically with increasing disorder strength. It is worth mentioning that the valley-polarized SC in RTG appears at relatively high values of the displacement field \cite{rtg_chiral_sc_experiment}, where the wavefunctions are strongly $A_{1}$-sublattice polarized. Consequently, the non-monotonic regime in \figref{fig:kinetics_and_interplay}b could be relevant to experiment.

\subsection{Finite-momentum pairing}

Because of the energy mismatch $\xi_{\mathbf{k}} \neq \xi_{-\mathbf{k}}$ (or broken TRS in the wave functions \cite{QGDiscrepancy}), $\vec{q}=0$ in \equref{SCOrderParameter} might not necessarily be energetically favorable, and pairing at finite-momentum can emerge, increasing the particle-hole nesting and suppressing the kinetic pair-breaking effect of $\xi_{\mathbf{k}+\mathbf{q}/2} - \xi_{-\mathbf{k}+\mathbf{q}/2}$. It is, therefore, interesting to study the stability of such pairing in the presence of impurities. Although the discussion of quantum geometric pair breaking in the weak disorder limit is straightforwardly generalized to finite $\vec{q}$ (see SI), we find it convenient to use a numerical approach to obtain reliable results for the kinetic pair breaking effects in R$n$G. To focus on the latter effects, we investigate structureless, non-magnetic impurities $W_{\mathbf{k,\mathbf{k}'}} \to 1$. We further take $\Delta_{\mathbf{k}} = \Delta_0$ and dispersion of the full continuum model of R$4$G for concreteness.

\begin{figure}[t]
    \centering
    \includegraphics[width=0.9\linewidth]{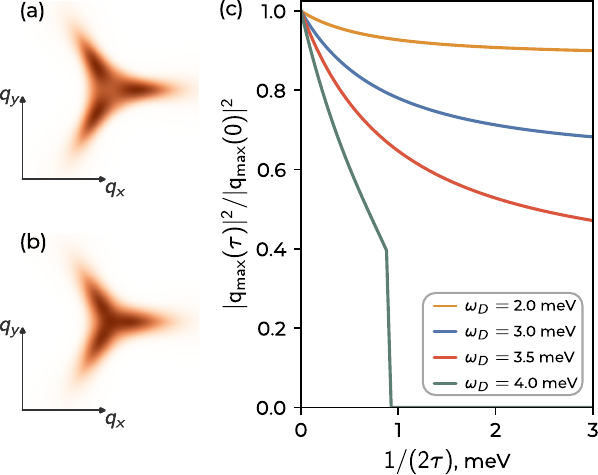}
    \caption{\textbf{Finite momentum pairing.} (a) Position of the maximum of the particle-particle bubble $\chi^{C}(\mathbf{q})$ as a function of the disorder strength for different values of the cutoff frequency $\omega_D$. For $\omega_D = 4 \text{ meV}$, the 2D map of $\chi^{C}(\mathbf{q})$ is shown in the clean limit (b) and disordered limit (c) when the finite-momentum pairing is fully suppressed.}
    \label{fig:finite_momentum}
\end{figure}

In \figref{fig:finite_momentum}a-b, we present the profile of the clean and disordered particle-particle bubble $\chi^{C}(\mathbf{q})$ as a function of $\mathbf{q}$, revealing a suppression of its maximum $\mathbf{q}_{\mathrm{max}}$.
This can be seen more systematically in \figref{fig:finite_momentum}c, where we directly show how $\mathbf{q}_{\mathrm{max}}$ depends on the disorder strength for different values of the cutoff frequency. We find that, independent of the specific values of $\omega_D$, $\mathbf{q}_{\mathrm{max}}$ is suppressed by  disorder. These results are intuitive, as we saw above that scattering off impurities averages out the antisymmetric part of the energy, which is an important component of the formation of finite-momentum pairing. 

Whether sufficiently strong disorder eventually leads to a full suppression of $\mathbf{q}_{\mathrm{max}} \rightarrow 0$ depends on the cutoff frequency. Namely, for $\omega_{D} = 2, 3, 3.5 \text{ meV}$ we find that disorder monotonically suppresses $\mathbf{q}_{\mathrm{max}}$, pinning it at a finite value $\mathbf{q}_{\mathrm{max}}(\tau^{-1} \to \infty) \neq 0$ in the limit of infinitely strong disorder. For $\omega_D = 4 \text{ meV}$, however, we find that it monotonically suppresses $\mathbf{q}_{\mathrm{max}}$, but only until some critical value of $\tau^{-1}$, at which point $\mathbf{q}_{\mathrm{max}}(\tau^{-1})$ discontinuously vanishes in a first order transition.

\section{Discussion}
In this work, we have analyzed the impact of impurity scattering on the stability of superconductivity emerging from a normal-state with broken time-reversal symmetry, i.e., without any anti-unitary relation between the Bloch states $\ket{\phi_{\vec{k}}}$ and $\ket{\phi_{-\vec{k}}}$. Both wave-function effects as well as kinetic pair breaking have been taken into account and are shown to be essential for a complete description. These two effects, their consequences, and interplay can be most compactly seen in our general relation of the dimensionless disorder sensitivity $\zeta$ in \equref{eq:general_zeta}, where they enter in the form of the disorder matrix elements $W_{\vec{k},\vec{k}'} = \braket{\phi_{\vec{k}}|w|\phi_{\vec{k}'}}$ and the antisymmetric part $a_{\vec{k}} = ( \xi_{\vec{k}}- \xi_{-\vec{k}} ) / 2$ of the normal-state dispersion, respectively. 

First, our findings connect quantum geometry, which has previously been very actively studied in the context of different aspects of superconductivity, and the disorder sensitivity of the pairing state. We found that quantum geometry can lead to pair breaking in any superconducting state which we connect back to its obstruction in maximizing the weight of the Cooper-pair wavefunction in \equref{eq:Dx_def} in a single unit cell with respect to the measure in \equref{eq:Cx_def}. 
As this relation is very general and there are many systems with non-trivial quantum geometry and a rich set of impurities to study, this directly opens up a very broad set of possible follow-up studies. It will also be interesting to see whether other impurity-related features in superconductors, such as local density of states effects, can also be directly related to quantum geometry. 

Second, we have seen that significant chirality in the normal state Bloch states can lead to a chiral topological or triplet superconductor being most stable against certain types of nominally non-magnetic impurities. Within the localization picture above, this can be seen as a result of the $\vec{k}$-dependent phases in the Bloch states (normal-state chirality) leading to maximal localization of $D(\vec{x})$ in \equref{eq:Dx_def} for a non-trivial winding of the superconducting order parameter. This also leads to the counterintuitive result that if multiple pairing states, including a trivial (or spin singlet) superconductor and a topological chiral (triplet) state, are energetically close, impurities can be beneficial for realizing a chiral topological state (triplet pairing).

Third, by explicit computations for rhombohedral graphene, we showed that these quantum geometric effects lead to a significant dependence of the disorder-sensitivity of the critical temperature on the layer and type of pairing state. As such, using controlled disorder in this and other related systems, such as twisted MoTe$_2$, could be used as a versatile tool to pinpoint the microscopic form of the pairing state. This is further corroborated by the fact that we showed that the kinetic pair breaking effects can lead to a non-monotonic dependence of $T_c$ on the disorder concentration, including an initial increase of it, see \figref{fig:kinetics_and_interplay}. If observed in these systems, this could be seen as direct evidence that superconductivity and valley polarization coexist microscopically.  

Finally, we showed that these aspects are also directly relevant to superconductivity coexisting with or proximitized by altermagnetism, where quantum geometry plays a dual role: it leads to pair breaking for weak altermagnetism but allows for superconducting states with $\zeta < 1/2$ in \equref{DimensionlessSuppressionOfTc}. Therefore, quantum geometric pair breaking also plays an important role in the emerging field of superconducting altermagnets. 

\begin{acknowledgments}
All authors acknowledge funding by the European Union (ERC-2021-STG, Project 101040651---SuperCorr). Views and opinions expressed are however those of the authors only and do not necessarily reflect those of the European Union or the European Research Council Executive Agency. Neither the European Union nor the granting authority can be held responsible for them. We thank Sayan Banerjee, Elio K\"onig, Ivan Iorsh, and J\"org Schmalian for discussions. 
\end{acknowledgments}

\bibliography{draft_Refs}

\onecolumngrid

\appendix

\section{Derivation of the general results for the particle-particle bubble}

\begin{figure}[h]
    \centering
    \includegraphics[width=0.4\linewidth]{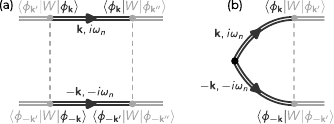}
    \caption{Blocks entering the disordered particle-particle bubble.}
    \label{fig:diagrams_SI}
\end{figure}

As discussed in the main text, in our calculations, we split the energy into the symmetric and antisymmetric parts, $\xi_{\mathbf{k}} =  \tilde{\xi}_{\mathbf{k}} + a_{\mathbf{k}}, \tilde{\xi}_{\mathbf{k}} = ( \xi_{\mathbf{k}} + \xi_{-\mathbf{k}} )/2,  a_{\mathbf{k}} = ( \xi_{\mathbf{k}} - \xi_{-\mathbf{k}} )/2$, and rewrite all summations over $\mathbf{k}$ using the following generalized constant-DOS approximation,
\begin{align}
    \sum_{\mathbf{k}} \ldots \to \int d\varphi_{\mathbf{k}} \tilde{\rho}(\varphi_{\mathbf{k}}) \int d\tilde{\xi} \ldots,
\end{align}
where $\tilde{\rho}(\varphi_{\mathbf{k}})$ is the angular-resolved density of states of the symmetrized energies $\tilde{\xi}_{\mathbf{k}}$ at the associated symmetrized Fermi surface $\tilde{\xi}_{\mathbf{k}} = 0$, the integral over $\tilde{\xi}$ corresponds to summation over the momentum perpendicular to the above-mentioned Fermi surface. We will also assume that $a_{\mathbf{k}}$ and the disorder matrix elements $W_{\mathbf{k},\mathbf{k}'}$ depend only on the angle $\varphi_{\mathbf{k}/\mathbf{k}'}$ and not on $\tilde{\xi}$.

We first calculate the perturbative self-energy,
\begin{align}
    \Sigma(i\omega_n, \mathbf{k}) = \gamma^2 \sum_{\mathbf{k}'} |W_{\mathbf{k}\mathbf{k}'}|^2 G_0(i\omega_n, \mathbf{k}') = -i  \sign(\omega_n) \pi \gamma^2 \int d  \varphi_{\mathbf{k}'} \tilde{\rho}(\varphi_{\mathbf{k}'}) | W_{\mathbf{k}\mathbf{k}'} |^{2} = -i \sign(\omega_n) \Gamma_{\mathbf{k}},
\end{align}
where we used the fact that $\int d\tilde{\xi} (i\omega_n - \tilde{\xi} - a_{\mathbf{k}})^{-1} = -i \pi \sign(\omega_n)$. Because of the trivial identity
\begin{align}
    \int d\xi G_{0}^{}(i\omega_n \to i\omega_n + i\sign(\omega_n)\Gamma, \xi) = \int d\xi G_{0}(i\omega_n, \xi),
\end{align}
the obtained self-energy also corresponds to the full self-consistent non-crossing result. Thus, the disordered Green's function reads
\begin{align}
    G^{-1}(i\omega_n,\mathbf{k}) = G^{-1}_0(i\omega_n,\mathbf{k}) - \Sigma(\omega_n,\mathbf{k}) = i\omega_n + i\sign(\omega_n) \Gamma_{\mathbf{k}} - \xi_{\mathbf{k}}.
\end{align}

Next we focus on calculating the building block of the vertex corrections which is diagramatically presented in Fig.~\ref{fig:diagrams_SI}a (grey lines do not enter the definition of the block, and they are presented for convenience),
\begin{align}
    B(i\omega_n) = \sum_{\mathbf{k}} G(i\omega_n,\mathbf{k}) G(-i\omega_n,-\mathbf{k}) \ket{\phi_{\mathbf{k}}}\bra{\phi_\mathbf{k}} \otimes \ket{\phi_{-\mathbf{k}}}\bra{\phi_{-\mathbf{k}}},
\end{align}
where the direct product is taken between the spaces associated with upper and lower fermionic lines. The integral over $\tilde{\xi}$ yields
\begin{align}
    \int d\tilde{\xi} \frac{1}{i\omega_n + i\sign(\omega_n)\Gamma_k - \tilde{\xi} - a_{\mathbf{k}}} \frac{1}{-i\omega_n - i\sign(\omega_n)\Gamma_k - \tilde{\xi} + a_{\mathbf{k}}} = \pi \frac{|\omega_n| + (\Gamma_{\mathbf{k}} + \Gamma_{-\mathbf{k}})/2  }{ (|\omega_n| + (\Gamma_{\mathbf{k}} + \Gamma_{-\mathbf{k}})/2)^2 + a^2_{\mathbf{k}} },
\end{align}
hence,
\begin{align}
    B(i\omega_n) = \pi \int d\varphi_{\mathbf{k}} \tilde{\rho}(\varphi_{\mathbf{k}}) \frac{|\omega_n| + (\Gamma_{\mathbf{k}} + \Gamma_{-\mathbf{k}})/2  }{ (|\omega_n| + (\Gamma_{\mathbf{k}} + \Gamma_{-\mathbf{k}})/2)^2 + a^2_{\mathbf{k}} } \ket{\phi_{\mathbf{k}}}\bra{\phi_\mathbf{k}} \otimes \ket{\phi_{-\mathbf{k}}}\bra{\phi_{-\mathbf{k}}}.
\end{align}
It is convenient to also define the edge block of the particle-particle bubble shown in Fig.~\ref{fig:diagrams_SI}b,
\begin{align}
    B_{e}^\dagger (i\omega_n) = \pi \int d\varphi_{\mathbf{k}} \tilde{\rho}(\varphi_{\mathbf{k}}) \frac{|\omega_n| + (\Gamma_{\mathbf{k}} + \Gamma_{-\mathbf{k}})/2  }{ (|\omega_n| + (\Gamma_{\mathbf{k}} + \Gamma_{-\mathbf{k}})/2)^2 + a^2_{\mathbf{k}} } \bra{\phi_{\mathbf{k}}} \otimes \bra{\phi_{-\mathbf{k}}}.
\end{align}
Thus, full disordered Cooperon ladder can be written in terms of defined blocks as follows
\begin{align}
    \Pi^{C}(i\omega_n) = \pi \int d\varphi_{\mathbf{k}} \tilde{\rho}(\varphi_{\mathbf{k}}) \frac{|\omega_n| + (\Gamma_{\mathbf{k}} + \Gamma_{-\mathbf{k}})/2  }{ (|\omega_n| + (\Gamma_{\mathbf{k}} + \Gamma_{-\mathbf{k}})/2)^2 + a^2_{\mathbf{k}} } + B_e^\dagger \gamma^2 w \otimes w \left( \mathds{1} - B\gamma^2 w \otimes w \right)^{-1} B_e,
\end{align}
where the first part correspond to a particle-particle bubble constructed with dressed Green's function and bare vertex, and the second term is the resummed series of ladder diagrams which include at least one disorder rung.

Thus, for the constant SC order parameter $\Delta_{\mathbf{k}} = \Delta_0$, the quadratic part of the Ginzburg-Landau theory reads
\begin{align}
    \braket{\mathcal{F}}_{\mathrm{dis}} \sim \Delta_0^* \left( - T\sum_{\omega_n} \Pi^{C}(i\omega_n) + \frac{1}{g} \right) \Delta_0,
\end{align}
where $g$ is an effective interaction constant.

It is straightforward to derive the generalization of this expression for an arbitrary $\Delta_{\mathbf{k}}$. Let us first define the $\mathbf{k}$-dependent version of the edge block,
\begin{align}
    B_{e}^\dagger(i\omega_n,\mathbf{k}) = \pi \frac{|\omega_n| + (\Gamma_{\mathbf{k}} + \Gamma_{-\mathbf{k}})/2  }{ (|\omega_n| + (\Gamma_{\mathbf{k}} + \Gamma_{-\mathbf{k}})/2)^2 + a^2_{\mathbf{k}} } \bra{\phi_{\mathbf{k}}} \otimes \bra{\phi_{-\mathbf{k}}},
\end{align}
and for convenience, the summation over the symmetrized Fermi surface as follows,
\begin{align}
    \sum_{\mathbf{k}}^{\mathrm{FS}} = \frac{1}{\tilde{\nu}} \int d\varphi_{\mathbf{k}}  \tilde{\rho}(\varphi_{\vec{k}})
\end{align}
with $\tilde{\nu} = \int d\varphi_{\mathbf{k}} \tilde{\rho}(\varphi_{\mathbf{k}})$. Then the quadratic part is given by the following expression
\begin{align}
    \braket{\mathcal{F}}_{\mathrm{dis}} \sim \sum_{\mathbf{k},\mathbf{k}'}^{\mathrm{FS}} \Delta_{\mathbf{k}}^* \left[ - \chi^{C}_{\mathbf{k},\mathbf{k}'}(T) + (g^{-1})_{\mathbf{k},\mathbf{k}'} \right] \Delta_{\mathbf{k}'} \label{eq:GL_general_SI}
\end{align}
where $g$ is now the kernel of the interaction mediating the SC, and the particle-particle bubble is given by
\begin{align}
    \chi^{C}_{\mathbf{k},\mathbf{k}'}(T) = T \sum_{\omega_n} \left( \pi \frac{|\omega_n| + (\Gamma_{\mathbf{k}} + \Gamma_{-\mathbf{k}})/2  }{ (|\omega_n| + (\Gamma_{\mathbf{k}} + \Gamma_{-\mathbf{k}})/2)^2 + a^2_{\mathbf{k}} } \delta_{\mathbf{k},\mathbf{k}'} +  B_e^\dagger(i\omega_n,\mathbf{k}) \gamma^2 w \otimes w \left\{ \mathds{1} - B(i\omega_n)\gamma^2 w \otimes w \right\}^{-1} B_e(i\omega_n,\mathbf{k}') \right).
\end{align}
Although the particle-particle bubble can be written in the more compact form, we find this expression to be particularly useful for numerical calculations as it only involves the inversion of the matrix $ \mathds{1} - B(i\omega_n)\gamma^2 w \otimes w $, whose dimensionality is determined by the number of internal degrees of freedom.

\section{Weak disorder expansion}

\begin{figure}[b]
    \centering
    \includegraphics[width=0.5\linewidth]{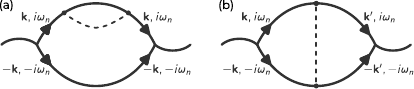}
    \caption{Diagrams corresponding to the first order disorder corrections to the particle-particle bubble.}
    \label{fig:perturbative_diagrams_si}
\end{figure}

We now derive the generalized version of the Abrokosov-Gorkov law in the SC emerging from a TRS-broken band. To do this, we first expand $\chi^{C}$ to the leading order of $\gamma^2$,
\begin{align}
    \chi^{C}_{\mathbf{k},\mathbf{k}'}(T) = \pi T \sum_{\omega_n} \left( \frac{|\omega_n| }{ \omega_n^2 + a^2_{\mathbf{k}} } \delta_{\mathbf{k},\mathbf{k}'} + \frac{\Gamma_{\mathbf{k}} + \Gamma_{-\mathbf{k}} }{2} \frac{a_{\mathbf{k}}^2 - \omega_n^2}{ (\omega_n^2 + a^2_{\mathbf{k}})^2 } \delta_{\mathbf{k},\mathbf{k}'} + \pi \gamma^2 \frac{|\omega_n| }{ \omega_n^2 + a^2_{\mathbf{k}} } \frac{|\omega_n| }{ \omega_n^2 + a^2_{\mathbf{k}'} } W_{\mathbf{k},\mathbf{k}'} W_{-\mathbf{k},-\mathbf{k}'} \right) + o(\gamma^2).
\end{align}
The summation over Matsubara frequency yields
\begin{align}
    \pi T \sum_{\omega_n} \frac{ |\omega_n| }{\omega_n^2 + a_{\vec{k}}^2} &= \sum_{n \geq 0} \frac{n+1/2}{(n+1/2)^2 + (a_{\mathbf{k}}/(2\pi T))^2} \nonumber \\
    &= \frac{1}{2} \sum_{n \geq 0} \left[ \frac{1}{n + 1/2 + ia_{\mathbf{k}}/(2\pi T)} + \frac{1}{n + 1/2 - ia_{\mathbf{k}}/(2\pi T)} \right] \nonumber \\
    &= - \frac{1}{2} \sum_{n \geq 0} \left[ \frac{1}{n + 1} -  \frac{1}{n + 1/2 + ia_{\mathbf{k}}/(2\pi T)} + \frac{1}{n + 1} -  \frac{1}{n + 1/2 - ia_{\mathbf{k}}/(2\pi T)} \right] + \sum_{n\geq 0} \frac{1}{n+1} \nonumber \\
    &= \log \left( \frac{\omega_D}{2\pi T} \right) - \frac{1}{2} \psi\left( \frac{1}{2} + i \frac{a_{\mathbf{k}}}{2\pi T} \right) - \frac{1}{2} \psi\left( \frac{1}{2} - i \frac{a_{\mathbf{k}}}{2\pi T} \right) \eqcolon f_0(\mathbf{k},T),\\
    T \sum_{\omega_n} \frac{a_{\vec{k}}^2 - \omega_n^2}{(\omega_n^2 + a_{\vec{k}}^2)^2} &= - \frac{1}{4 T} \cosh^{-2} \left( \frac{a_\vec{k}}{2T} \right) \eqcolon - \frac{1}{2T} f_1^2(\mathbf{k}, T),\\
    T \sum_{\omega_n} \frac{\omega_n^2}{(\omega_n^2 + a_{\vec{k}}^2)(\omega_n^2 + a_{\vec{k}'}^2)} &= \frac{a_{\vec{k}} \tanh[a_{\vec{k}}/(2 T)] - a_{\vec{k}'} \tanh[a_{\vec{k}'}/(2 T)]}{2(a_{\vec{k}}^2 - a_{\vec{k}'}^2)} \eqcolon \frac{1}{2T} f_2^2(\mathbf{k}, \mathbf{k}', T),
\end{align}
where $\psi(z)$ is the digamma function and $\omega_D$ is the cutoff frequency regularizing logarithmic a divergence of the bare particle-particle bubble. One can think of the introduced functions $f_{1}(\mathbf{k}, T), f_{2}(\mathbf{k}, \mathbf{k}', T)$ as spectral weights of the diagrams corresponding to the first order correction to the particle-particle bubble presented in \figref{fig:perturbative_diagrams_si}, respectively. 

To find the disordered critical temperature $T_{c}$ we should also expand bare particle-particle bubble near the clean $T_{c,0}$,
\begin{align}
   f_{0}(\mathbf{k},T_{c,0} + (T_{c} - T_{c,0})) &= f_{0}(\mathbf{k},T_{c,0}) - f_{3}(\mathbf{k},T_{c,0}) \frac{T_{c} - T_{c,0}}{T_{c,0}} + o \left(\frac{T_{c} - T_{c,0}}{T_{c,0}}\right),\\
    f_{3}(\mathbf{k},T_{c,0}) & \coloneq 1 + \frac{i a_{\mathbf{k}}}{2\pi T_{c,0}} \left[ \psi'\left( \frac{1}{2} + i \frac{a_{\mathbf{k}}}{2\pi T_{c,0}} \right) - \psi'\left( \frac{1}{2} -i \frac{a_{\mathbf{k}}}{2\pi T_{c,0}} \right) \right] .
\end{align}

We now rewrite the summation over momenta in terms of the following weighted commutators,
\begin{align}
    C^{\pm}_{\mathbf{k},\mathbf{k}',f_1}(T) &= W_{\mathbf{k}\mathbf{k}'} \Delta_{\mathbf{k}'} f_{1}(\mathbf{k}',T) \pm f_{1}(\mathbf{k},T) \Delta_{\mathbf{k}} W_{-\mathbf{k},-\mathbf{k}'}^{*},\\
    C^{\pm}_{\mathbf{k},\mathbf{k}',f_2}(T) &= W_{\mathbf{k}\mathbf{k}'} f_{2}(\mathbf{k},\mathbf{k}',T) \Delta_{\mathbf{k}'}  \pm f_{2}(-\mathbf{k},-\mathbf{k}',T) \Delta_{\mathbf{k}} W_{-\mathbf{k},-\mathbf{k}'}^{*},\\
    C^{\pm}_{\mathbf{k},\mathbf{k}'} &= W_{\mathbf{k}\mathbf{k}'} \Delta_{\mathbf{k}'}  \pm  \Delta_{\mathbf{k}} W_{-\mathbf{k},-\mathbf{k}'}^{*}.
\end{align}

\newpage

\begin{align}
    \frac{1}{2} \sum_{\mathbf{k}}^{\mathrm{FS}} |\Delta_{\mathbf{k}}|^2 f_1^2(\mathbf{k},T) (\Gamma_{\mathbf{k}} + \Gamma_{\mathbf{k}}) &= \frac{\pi\gamma^2\tilde{\nu}}{2} \sum_{\mathbf{k},\mathbf{k}'}^{\mathrm{FS}} |\Delta_{\mathbf{k}}|^2 f_1^2(\mathbf{k},T) \left( |W_{\mathbf{k},\mathbf{k}'}|^2 + |W_{-\mathbf{k},\mathbf{k}'}|^2 \right) \nonumber \\
    &= \frac{\pi\gamma^2\tilde{\nu}}{2} \sum_{\mathbf{k},\mathbf{k}'}^{\mathrm{FS}} |\Delta_{\mathbf{k}}|^2 f_1^2(\mathbf{k},T) \left( |W_{\mathbf{k},\mathbf{k}'}|^2 + |W_{-\mathbf{k},-\mathbf{k}'}|^2 \right) \nonumber \\
    &= \frac{\pi\gamma^2\tilde{\nu}}{2} \sum_{\mathbf{k},\mathbf{k}'}^{\mathrm{FS}} \left( |W_{\mathbf{k},\mathbf{k}'}|^2 |\Delta_{\mathbf{k}'}|^2 f_1^2(\mathbf{k}',T) + f_1^2(\mathbf{k},T) |\Delta_{\mathbf{k}}|^2 |W_{-\mathbf{k},-\mathbf{k}'}|^2 \right) \nonumber \\
    &= \frac{\pi\gamma^2\tilde{\nu}}{4} \sum_{\mathbf{k},\mathbf{k}'}^{\mathrm{FS}} \left( | C_{\mathbf{k},\mathbf{k}',f_1}^{+} |^2 + | C_{\mathbf{k},\mathbf{k}',f_1}^{-} |^2  \right),\\
    \sum_{\mathbf{k},\mathbf{k}'}^{\mathrm{FS}} \Delta_{\mathbf{k}}^* f_2^2(\mathbf{k},\mathbf{k}',T) W_{\mathbf{k},\mathbf{k}'} W_{-\mathbf{k},-\mathbf{k}'} \Delta_{\mathbf{k}'} &= \frac{1}{2} \sum_{\mathbf{k},\mathbf{k}'}^{\mathrm{FS}} f_2^2(\mathbf{k},\mathbf{k}',T) \left( \Delta^*_{\mathbf{k}} W_{\mathbf{k}\mathbf{k}'} W_{-\mathbf{k},-\mathbf{k}'} \Delta_{\mathbf{k}'} + \Delta^*_{\mathbf{k}'} W_{\mathbf{k}',\mathbf{k}} W_{-\mathbf{k}',-\mathbf{k}} \Delta_{\mathbf{k}}  \right) \nonumber \\
    &=\frac{1}{2} \sum_{\mathbf{k},\mathbf{k}'}^{\mathrm{FS}} f_2^2(\mathbf{k},\mathbf{k}',T) \left( \Delta^*_{\mathbf{k}'} W_{\mathbf{k}\mathbf{k}'} W_{-\mathbf{k},-\mathbf{k}'} \Delta_{\mathbf{k}} + \Delta^*_{\mathbf{k}'} W_{\mathbf{k}\mathbf{k}'}^* W_{-\mathbf{k},-\mathbf{k}'}^* \Delta_{\mathbf{k}}  \right) \nonumber \\
    &= \frac{1}{4} \sum_{\mathbf{k},\mathbf{k}'}^{\mathrm{FS}} \left( | C_{\mathbf{k},\mathbf{k}',f_2}^{+} (T) |^2 - | C_{\mathbf{k},\mathbf{k}',f_2}^{-}(T) |^2  \right).
\end{align}

Finally, we define the average lifetime as follows 
\begin{align}
    \frac{1}{2\tau} = \sum_{\mathbf{k}}^{\mathrm{FS}} \Gamma_{\mathbf{k}} = \pi \gamma^2 \tilde{\nu} \sum_{\mathbf{k},\mathbf{k}'}^{\mathrm{FS}} |W_{\mathbf{k},\mathbf{k}'}|^2,
\end{align}
and obtain the generalization of the Abrikosov-Gorkov law,
\begin{align}
    \frac{T_{c} - T_{c,0}}{T_{c,0}} = -\frac{\pi}{4 T_{c,0}} \tau^{-1} \zeta + o(\tau^{-1})
\end{align}
where
\begin{align}
    \zeta = \left( \sum_{t=\pm} \sum_{\mathbf{k},\mathbf{k}'}^{\mathrm{FS}} \left| C_{\mathbf{k},\mathbf{k}',f_1}^{t}(T_{c,0}) \right|^2 - t  \left| C_{\mathbf{k},\mathbf{k}',f_2}^{t}(T_{c,0}) \right|^2 \right)  \left( 4 \sum_{\mathbf{k},\mathbf{k}'}^{\mathrm{FS}}  \left| W_{\mathbf{k}\mathbf{k}'} \right|^2 \sum_{\mathbf{k}}^{\mathrm{FS}} f_3(\mathbf{k},T_{c,0}) |\Delta_{\mathbf{k}}|^2 \right)^{-1}.
\end{align}
Since the behavior of this expression is hard to analyze analytically, we simplify it by considering the small $a_{\mathbf{k}}$ limit, $\| a_{\mathbf{k}} \| / T_{c,0} \ll 1$. We first note that when $a_{\mathbf{k}} = 0$, $C_{\mathbf{k},\mathbf{k}',f_{1/2}}^{\pm}(T_{c,0}) \to 1/2 C_{\mathbf{k},\mathbf{k}'}^{\pm}$ and we recover the result from~\cite{PhysRevResearch.2.023140},
\begin{align}
    \zeta = \left( \sum_{\mathbf{k},\mathbf{k}'} | C_{\mathbf{k},\mathbf{k}'}^{-} |^2  \right) \left( 4 \sum_{\mathbf{k}}^{\mathrm{FS}} |\Delta_{\mathbf{k}}|^2 \sum_{\mathbf{k},\mathbf{k}'}^{\mathrm{FS}}  \left| W_{\mathbf{k}\mathbf{k}'} \right|^2 \right)^{-1}.
\end{align}
Including the leading order of $a_{\mathbf{k}}/T_{c,0}$, we get
\begin{align}
    \begin{aligned}
        \zeta = \sum_{\mathbf{k},\mathbf{k}'}^{\mathrm{FS}} &\left(  | C_{\mathbf{k},\mathbf{k}'}^{-} |^2 \left[ 1 - \frac{a_{\mathbf{k}}^2 + a_{\mathbf{k}'}^2}{24 T_{c,0}^2} + \frac{7 \zeta_{\mathrm{R}}(3)}{2\pi^2} \sum_{\mathbf{k}_1}^{\mathrm{FS}} \frac{a_{\mathbf{k}_1}^2}{T_{c,0}^2} \frac{|\Delta_{\mathbf{k}_1}|^2}{\sum_{\mathbf{k}_1'}^{\mathrm{FS}} |\Delta_{\mathbf{k}_1'}|^2 }  \right] \right. \\
        & \left. - |C_{\mathbf{k},\mathbf{k}'}^{+}|^2 \frac{a_{\mathbf{k}}^2 + a_{\mathbf{k}'}^2}{T_{c,0}^2} - \frac{1}{8} \sum_{t=\pm} \left| W_{\mathbf{k}\mathbf{k}'} \Delta_{\mathbf{k}'} \frac{|a_{\mathbf{k'}}|}{T_{c,0}} + t \frac{|a_{\mathbf{k}}|}{T_{c,0}} \Delta_{\mathbf{k}} W_{-\mathbf{k},-\mathbf{k}'}^{*} \right|^2 \right) \left( 4 \sum_{\mathbf{k}}^{\mathrm{FS}} |\Delta_{\mathbf{k}}|^2 \sum_{\mathbf{k},\mathbf{k}'}^{\mathrm{FS}}  \left| W_{\mathbf{k}\mathbf{k}'} \right|^2 \right)^{-1} 
    \end{aligned}
\end{align}
where $\zeta_{\mathrm{R}}(z)$ is the Riemann $\zeta$-function. 

It is instructive to consider the case of the SC order parameter with constant amplitude $|\Delta_{\mathbf{k}}| = \Delta_0$, and choose a gauge in which it is constant, then the disorder sensitivity factor $\zeta$ can be split into two terms, $\zeta = \zeta^+ + \zeta^-$,
\begin{align}
    \zeta^{+} & = - \left( \frac{1}{48} \sum_{\mathbf{k},\mathbf{k}'}^{\mathrm{FS}} \left| W_{\mathbf{k}\mathbf{k}'}^{+} \right|^2 \frac{a_{\vec{k}}^2 + a_{\vec{k}'}^2}{T_{c,0}^2} \right) \left( \sum_{\mathbf{k},\mathbf{k}'}^{\mathrm{FS}}  \left| W_{\mathbf{k}\mathbf{k}'} \right|^2 \right)^{-1} + o(\| a_{\mathbf{k}} \|^2),\\
    \zeta^{-} &= \left( \sum_{\mathbf{k},\mathbf{k}'}^{\mathrm{FS}} \left| W_{\mathbf{k}\mathbf{k}'}^{-} \right|^2 \left[ 1 - \frac{5}{48} \frac{a_{\vec{k}}^2 + a_{\vec{k}'}^2}{T_{c,0}^2} + \frac{7 \zeta_{\mathrm{R}}(3)}{2\pi^2} \sum_{\mathbf{k}_1}^{\mathrm{FS}} \frac{a_{\mathbf{k}_1}^2}{T_{c,0}^2} \right]  \right) \left( \sum_{\mathbf{k},\mathbf{k}'}^{\mathrm{FS}}  \left| W_{\mathbf{k}\mathbf{k}'} \right|^2  \right)^{-1} + o(\| a_{\mathbf{k}} \|^2).
\end{align}
where $W^{\pm}_{\mathbf{k}\mathbf{k}'}$ is time-reversal even and odd part of the disorder $W_{\mathbf{k},\mathbf{k}'}$. Here, we can clearly see that $\zeta^{+} < 0$ meaning that the nonmagnetic part of disorder can increase the critical temperature.

\section{Some comments on finite momentum pairing calculations}

Let us first highlight an artifact of the constant DOS approximation applied to the finite-$\mathbf{q}$ pairing. To simplify our calculations we only consider the simplest pairing, $\Delta_{0} d^\dagger_{\mathbf{k}+\mathbf{q}/2,\uparrow} d^\dagger_{-\mathbf{k}+\mathbf{q}/2,\downarrow}$, and nonmagnetic structureless disorder, $W_{\mathbf{k},\mathbf{k}'} \to 1$. It is tempting to introduce the generalized symmetric and antisymmetric part of the shifted energies, $\tilde{\xi}_{\mathbf{k},{\mathbf{q}}} =  \left( \xi_{\mathbf{k}+\mathbf{q}/2} + \xi_{-\mathbf{k}+\mathbf{q}/2} \right)/2, a_{\mathbf{k},\mathbf{q}} = \left(  \xi_{\mathbf{k}+\mathbf{q}/2} - \xi_{-\mathbf{k}+\mathbf{q}/2} \right)/2$, and then use a similar approximation as in the zero-$\mathbf{q}$ case,
\begin{align}
    \sum_{\mathbf{k}} \to \int_{\text{F.S.}(\tilde{\xi}_{\mathbf{k},\mathbf{q}})} d\varphi_{\mathbf{k}} \tilde{\rho}_{\mathbf{q}}(\varphi_{\mathbf{k}}) \int d\tilde{\xi}, \label{eq:finite_q_DOS_approx_SI}
\end{align}
where $\tilde{\rho}_{\mathbf{q}}(\varphi_{\mathbf{k}})$ is the DOS of the symmetrized energy $\left( \xi_{\mathbf{k}+\mathbf{q}/2} + \xi_{-\mathbf{k}+\mathbf{q}/2} \right)/2$. We then can use this approximation to calculate the block of the ladder diagram,
\begin{align}
    B(i\omega_n,\mathbf{q}) &= \sum_{\mathbf{k}} G(i\omega_n, \mathbf{k}+\mathbf{q}/2) G(-i\omega_n, -\mathbf{k}+\mathbf{q}/2) \label{RhoTildeBexpr}
\end{align}
For the self-energy calculation, however, it might no longer be justified to use the same approximation of the constant DOS~\eqref{eq:finite_q_DOS_approx_SI}, since the scattering rate should not depend on the momentum of the Cooper pair. Arguably, one should use the approximation that assumes that the DOS of $\xi_{\mathbf{k}}$ (denoted by $\rho(\varphi_{\mathbf{k}})$) is constant,
\begin{align}
    \Sigma(i\omega_n, \mathbf{k}) = \gamma^2 \sum_{\mathbf{k}} G_{0}(i\omega_n, \mathbf{k}) \to \gamma^2 \int d\xi \frac{1}{i\omega - \xi} \int d\varphi_{\mathbf{k}} \rho(\varphi_{\mathbf{k}}) = - i \sign(\omega_n) \pi \nu \gamma^2,
\end{align}
where $\nu = \int d\varphi_{\mathbf{k}} \rho(\varphi_{\mathbf{k}})$. From \equref{RhoTildeBexpr}, the block of the ladder can now be explicitly written as follows
\begin{align}
    B(i\omega_n,\mathbf{q}) = \pi \int d\varphi_{\mathbf{k}} \tilde{\rho}_{\mathbf{q}}(\varphi_{\mathbf{k}}) \frac{|\omega_n| + \pi \nu \gamma^2}{ (|\omega_n| + \pi \nu \gamma^2)^2 + a^2_{\mathbf{k},\mathbf{q}} }.
\end{align}
In the limit $\gamma \to \infty$, we get $\gamma^2 B(i\omega_n,\mathbf{q}) \to \tilde{\nu}_{\mathbf{q}} / \nu $ with $\tilde{\nu}_{\mathbf{q}} = \int d\varphi_{\mathbf{k}} \tilde{\rho}_{\mathbf{q}}(\varphi_{\mathbf{k}})$. In our numerical calculations for R$n$G, we find that $\tilde{\nu}_{\mathbf{q}} / \nu $ can be greater than one, and therefore, the particle-particle bubble, given by 
\begin{align}
    \Pi^{C}(i\omega_n, \mathbf{q}) = B(i\omega_n,\mathbf{q})\sum_{m=0}^{\infty} [\gamma^2 B(i\omega_n,\mathbf{q})]^m,
\end{align}
formally diverges when $\gamma \to \infty$. This is an artifact of the approximation suggesting that there is no straightforward way to generalize our analytical calculations to the disordered finite-momentum particle-particle bubble capturing both kinetic and geometric effects in all orders of $\gamma^2$. However, the isolated pair-breaking effect of quantum geometry in the weak disorder limit is still well-defined as shown in \appref{app:qg_finite_q}.

Thus, we focus on the numerical calculation of the finite-$\vec{q}$ bubble. We have found that its quantitative behavior depends a lot on the specific value of the cutoff frequency used to calculate the summation over $\mathbf{k}$. To make our numerical calculations consistent, we construct them using the Anderson theorem as an anchor. 

Let us consider the isolated TRS-broken band of interest and artificially create its full time-reversal copy. We then consider $\mathbf{k}$-independent pairing of full TRS partners. In this case, the clean particle-particle bubble reads
\begin{align}
    \Pi_{0,\text{TRS}}^{C}(i\omega_n) = \sum_{\mathbf{k},|\xi_{\mathbf{k}}|<\omega_D} \left| G_0(i\omega_n,\mathbf{k}) \right|^2 = \sum_{\mathbf{k},|\xi_{\mathbf{k}}|<\omega_D} \frac{1}{\omega_n^2 + \xi_{\mathbf{k}}^2}.
\end{align}
Disregarding the real part of the self-energy, the disordered Green's function is given by $G(i\omega_n,\mathbf{k}) = \left(i\omega_n - \xi_{\mathbf{k}} + i\sign(\omega_n) \Gamma(\omega_n) \right)^{-1}$ (the dependence of $\Gamma(\omega_n)$ on $\omega_n$ is kept), the one block of the ladder determining the vertex correction is given by
\begin{align}
    B_{\text{TRS}}(i\omega_n) = \sum_{\mathbf{k},|\xi_{\mathbf{k}}|<\omega_D} \left| G(i\omega_n,\mathbf{k}) \right|^2.
\end{align}
The Anderson theorem is fulfilled if the disordered particle-particle bubble is equal to the bare one, hence, after summing up all the ladder diagrams, we get the following condition
\begin{align}
    \Pi_{0,\text{TRS}}^{C}(i\omega_n) = \Pi_{\text{TRS}}^{C}(i\omega_n) = \frac{B_{\text{TRS}}(i\omega_n)}{1 - \gamma^2 B_{\text{TRS}}(i\omega_n)}.
\end{align}
We solve this nonlinear equation to find the scattering rate $\Gamma(i\omega_n)$. This is an important step, since the naive computation of the self energy with some cutoff frequency does not necessarily guarantee the fulfillment of the Anderson theorem. We then use the obtained scattering rate to compute the finite-momentum particle-particle bubble for the isolated TRS broken band that we started with,
\begin{align}
    \sum_{\mathbf{k}, |\xi_{\mathbf{k}+\mathbf{q}/2}| < \omega_D, |\xi_{\mathbf{k}+\mathbf{q}/2}| < \omega_D} G(i\omega_n, \mathbf{k}+\mathbf{q}/2) G(-i\omega_n, -\mathbf{k}+\mathbf{q}/2).
\end{align}
Note that here one should replace $\Gamma(i\omega_n) \to \Gamma(i\omega_n)/2$ as we reduce the number of bands back to one.

\section{Explicit form of the Cooper pair wave function}

Let us explicitly calculate the real space Cooper pair wavefunction for the minimal model of the rhombohedral n-layer graphene. Starting from the Hamiltonian
\begin{equation}
    h_{\vec{k}} =
    \begin{pmatrix}
        u_0 - \mu & t_0 (k_x - i k_y)^n \\ t_0 (k_x + i k_y)^n & -u_0 - \mu
    \end{pmatrix},
\end{equation}
we parametrize the Bloch wavefunction of the upper band at the Fermi surface as follows $\phi_{\mathbf{k}} = (\phi_{A}, \phi_{B} e^{in\varphi_{\mathbf{k}}})^T$ with $\phi_{A},\phi_B \in \mathds{R}$. The Cooper pair wavefunction is given by
\begin{align}
    D_{\alpha,\alpha'}(\vec{x}) = \sum_{\vec{k}} \Delta_{\vec{k}} \phi_{\vec{k},\alpha} \phi_{-\vec{k},\alpha'} e^{i\mathbf{kx}}, 
\end{align}
where $\Delta_{\mathbf{k}}$ includes projection onto the Fermi surface. By limiting our analysis to the special cases discusses in the main text, we replace $\Delta_{\mathbf{k}} \to \Delta_0 e^{im\varphi_{\mathbf{k}}} \delta(\tilde{\xi})/\tilde{\nu}, m\in\mathds{Z}$, which yields
 \begin{align}
    D(\mathbf{x}) = 2\pi \Delta_0 i^{m}
    \begin{pmatrix}
        J_{m}(k_{F} x) \phi_{A}^2 & i^{n} J_{n+m}(k_{F} x) \phi_{A}\phi_{B}\\
        i^{n} J_{n+m}(k_{F} x) \phi_{A}\phi_{B} & i^{2n} J_{2n+m}(k_{F} x) \phi_{B}^2
    \end{pmatrix},
\end{align}
where $J_{l}(z)$ is the Bessel function of the first kind.

Let us also calculate the disorder-induced measure of $D(\mathbf{x})$ for specific impurity matrix,
\begin{gather}
    C_{w}(\mathbf{x}) = \mathrm{Tr} \left[ D^\dagger(\vec{x}) w^* D(\vec{x}) w \right],\quad
    w = 
    \begin{pmatrix}
        w_{A} & 0\\
        0 & w_{B}
    \end{pmatrix},\\
    C_{w}(x) = (2\pi\Delta_0)^2 \left[ w_A^2 \phi_A^4 J_m^2(k_{F}x) + 2 w_{A} w_{B} \phi_A^2 \phi_B^2 J_{n+m}^2(k_{F}x) + w_{B}^2 \phi_B^4 J_{2n+m}^2(k_{F}x) \right]
\end{gather}

Considering for simplicity only positive $w_A, w_B$, we now rewrite this expression in the parametrization introduced in the main text, in which the matrix element of the impurity reads
\begin{gather}
    W_{\mathbf{k},\mathbf{k}'} = \phi_{\mathbf{k}}^\dagger w \phi_{\mathbf{k}'} = W_0\left[\eta^2 + (1 - \eta^2) e^{in(\varphi_{\mathbf{k}'} - \varphi_{\mathbf{k}})}\right],\\
    W_{0} = w_{A} \phi_{A}^2 + w_{B} \phi_{B}^2,\quad \eta^2 = \frac{w_{A} \phi_{A}^2}{w_{A} \phi_{A}^2 + w_{B} \phi_{B}^2}.
\end{gather}
In our calculation related to the sensitivty of the SC, $W_0$ naturally drops out, therefore, we just set it to $1$ and arrive at the following expression for $C_w(x)$,
\begin{align}
    C_{w}(x) = (2\pi\Delta_0)^2 \left[ \eta^4 J_m^2(k_{F}x) + 2 \eta^2 (1 - \eta^2) J_{n+m}^2(k_{F}x) + (1 - \eta^2)^2 J_{2n+m}^2(k_{F}x) \right].
\end{align}

\section{Phase conventions for strong spin-orbit coupling}\label{StrongSOC}
As the connection to the normal-state degrees of freedom is the most complex in case (2)---the limit of strong spin-orbit coupling---and to properly define it, we elaborate on it in this appendix. Just like in the main text, we start from a microscopic Hamiltonian, which we now directly state in momentum space,
\begin{equation}
    H = \sum_{\vec{k},\alpha,\beta} \psi_{\vec{k},\alpha}^\dagger (h_\vec{k})_{\alpha,\beta} \psi_{\vec{k},\beta}^\pdagger + \sum_{\vec{k}}\left[ \psi_{\vec{k},\alpha}^\dagger (\hat{\Delta}_\vec{k})_{\alpha,\beta} \psi_{-\vec{k},\beta}^\dagger  + \text{H.c.} \right] + H_{\mathrm{dis}}. \label{UnprojectkSpaceSCHam}
\end{equation}
Here $\psi_{\vec{k},\alpha}$ are related to $\psi_{\alpha}(\vec{r})$ by Fourier transform, and $\hat{\Delta}_\vec{k}$ denotes the superconducting order parameter matrix in the microscopic basis; $\alpha$ also includes spin as a quantum number, however, as opposed to case (3), we assume that there is strong, asymmetric spin-orbit coupling, such that, for the purpose of studying pairing, we can project onto the single, non-degenerate active band closest to the Fermi level for each momentum $\vec{k}$---this is referred to as the ``weak pairing limit'' in \refcite{DesignPrinciples}. Formally, the band projection corresponds to $\psi_{\vec{k},\alpha} \rightarrow (\phi_{\vec{k}})_{\alpha} d_{\vec{k}}$ and leads to 
\begin{equation}
    \Delta_{\vec{k}} = \sum_{\alpha,\beta} \phi_{\vec{k},\alpha}^*  (\hat{\Delta}_\vec{k})_{\alpha,\beta} \phi_{-\vec{k},\beta}^* \label{Delta1}
\end{equation}
in the first term in \equref{SCOrderParameter}. Naturally, $\hat{\Delta}_\vec{k} = -\hat{\Delta}_\vec{k}^T$ as imposed by Fermi Dirac statistics in \equref{UnprojectkSpaceSCHam}, automatically implying $\Delta_{\vec{k}} = -\Delta_{-\vec{k}}$.

In the special case with TRS in the normal state, we have $\ket{\phi_{-\vec{k}}} = e^{i\rho_{\vec{k}}}\Theta \ket{\phi_{\vec{k}}}$ where $e^{i\rho_{\vec{k}}} = - e^{i\rho_{-\vec{k}}}$ as a result of $\Theta^2 = -\mathbbm{1}$. Using this in \equref{Delta1}, we can write
\begin{equation}
    \Delta_{\vec{k}} = \widetilde{\Delta}_{\vec{k}} e^{-i\rho_{-\vec{k}}}, \quad \text{with} \quad \widetilde{\Delta}_{\vec{k}} = \braket{\phi_{\vec{k}}|\hat{\Delta}_\vec{k} T^\dagger|\phi_{\vec{k}}},
\end{equation}
where $T$ is the unitary part of the time-reversal operator $\Theta = T \mathcal{K}$ ($\mathcal{K}$ is complex conjugation). As shown in \refcite{DesignPrinciples}, $\widetilde{\Delta}_{\vec{k}} = \widetilde{\Delta}_{-\vec{k}}$ is a very natural representation of the superconductor order parameter as it transforms explicitly under the same irreducible representation as $\hat{\Delta}_\vec{k}$. For instance, $\widetilde{\Delta}_{\vec{k}} = \Delta_0$ transforms trivially under all point-group symmetries. 

We finally recall that we found $e^{i\bar{\alpha}_{\mathbf{k}}}=e^{-i(\rho_{\vec{k}}+\varphi_{\vec{k}})}$ in the main text as the general solution to the gauge-optimization problem starting with $\Delta_{\vec{k}} = \Delta_0 e^{i\varphi_{\vec{k}}}$. From the perspective of the gauge choice, this corresponds to choosing Bloch states, $\ket{\phi'_{\vec{k}}}$ which now obey $\ket{\phi'_{-\vec{k}}} = e^{-i\varphi_{\vec{k}}}\Theta \ket{\phi'_{\vec{k}}}$, i.e., effectively $e^{i\rho_{\vec{k}}} \rightarrow e^{-i\varphi_{\vec{k}}}$. 

Alternatively, one can this of this as changing the superconducting order parameter as follows:
\begin{equation}
    \Delta_{\vec{k}} \rightarrow \Delta_0 e^{i\varphi_{\vec{k}}}e^{-i(\rho_{\vec{k}}+\varphi_{\vec{k}})} = \Delta_0 e^{-i\rho_{\vec{k}}}, \quad \text{or, equivalently,} \quad \widetilde{\Delta}_{\vec{k}} \rightarrow -\Delta_0.
\end{equation}
This means that, in the presence of TRS, our gauge choice just corresponds to the superconductor with $|\Delta_{\vec{k}}| = \text{const.}$ that transforms trivially under all point group symmetries, naturally explaining why we find $\zeta_{\text{opt}}=0$ (Anderson theorem). Our analysis of the main text generalizes this phase choice to the case without TRS.

\section{Altermagnetic model}\label{AltermagneticModelDescr}
\begin{figure}[h]
    \centering
    \includegraphics[width=0.6\linewidth]{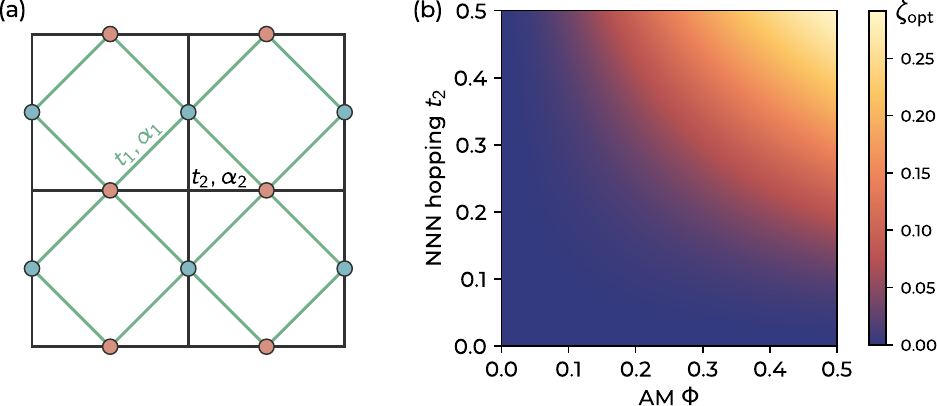}
    \caption{(a) Lattice representation of the Hamiltonian defined in momentum space in \equref{AMModelNS}. (b) Optimal $\zeta$ in the limit $\alpha_2=0$ at fixed $\alpha_1 = 0.05$, shown as a function of the altermagnetic order-parameter strength $\Phi$ and next-nearest-neighbor hopping $t_2$. Note that the system becomes an antiferromagnet in the limit $t_2 \rightarrow 0$, where $\zeta_{\text{opt}} \rightarrow 0$.}
    \label{fig:am_zeta_SI}
\end{figure}

To illustrate quantum geometric pair breaking in altermagnets, we use the same checkerboard-lattice model of itinerant electrons as in \refcite{PhysRevResearch.7.023152}, supplemented by spin-orbit coupling, see \figref{fig:am_zeta_SI}a. Denoting Pauli matrices in sublattice and spin space by $\tau_j$ and $s_j$, respectively, the normal-state Bloch Hamiltonian reads as

\begin{align}\begin{split}
    h_{\vec{k}} &= -2t_2s_0\left(P_+ \cos{k_x} +P_- \cos{k_y})\right) + t_1 (\tau_+ f_{\vec{k}} + \tau_- f^*_{\vec{k}}) + \Phi s_3 \tau_3  \\ &\qquad + \alpha_1 \left( i \sum_{p_1,p_2=\pm} \vec{v}_{p_1,p_2}\cdot\vec{s}\, \tau_+ e^{\frac{i}{2} (p_1 k_x+ p_2k_x)} +\text{H.c.} \right)  + \alpha_2 \left( s_2 P_+ \sin k_x - s_1 P_- \sin k_y \right), \label{AMModelNS}
\end{split}\end{align}
where $\tau_\pm = (\tau_x  \pm i \tau_y)/2$, $P_\pm = ( \tau_0 \pm \tau_3)/2$, $f_{\vec{k}} = \sum_{p_{1,2}=\pm }e^{\frac{i}{2} (p_1 k_x+ p_2k_x)} = 4\cos \frac{k_x}{2} \cos \frac{k_y}{2}$, and $\vec{v}_{+,+}=(-1,1)$, $\vec{v}_{+,-}=(1,1)$, $\vec{v}_{-,+}=(-1,-1)$, $\vec{v}_{-,-}=(1,-1)$. Physically, $t_j$ corresponds to $j$th-nearest neighbor hopping, $\Phi$ is the altermagnetic order parameter, and $\alpha_1$ ($\alpha_2$) is the strength of nearest-neighbor (next-nearest-neighbor) spin-orbit coupling. For concreteness, we set $\alpha_1=0$, $\alpha_2=\alpha$ in the main text. All energies are measured in units of $t_1$ and we set $t_2=0.3$.

Importantly, when $\alpha_j = 0$ and $\Phi\neq 0$, the system exhibits an ``orbital'' TRS, $\Theta = \mathcal{K}$ with $\Theta^2 = \mathbbm{1}$, while the usual spinful TRS ($\Theta = i s_2 \mathcal{K}$) is broken by the altermagnetic order parameter. The opposite is true in the limit $\alpha \neq 0$ and $\Phi = 0$, where the preserved TRS is spinful, $\Theta = i s_2 \mathcal{K}$ and obeys $\Theta^2 = -\mathbbm{1}$. 

Naturally, there is no TRS when both $\alpha$ and $\Phi$ are non-zero. Nonetheless, it holds $\xi_{\vec{k}} = \xi_{-\vec{k}}$ which follows from spinful $C_{2z}$ rotational symmetry, with representation $\hat{C}_{2z} =  i s_3 \tau_0$ within the conventions of \equref{AMModelNS}. As such, for zero-momentum pairing, there is no kinetic pair breaking effect. The magnetic point group of the model is generated by the product of four-rotation and time-reversal, $\Theta C_{4z}$, as well as the combination of reflection at the $xz$ plane and time-reversal, $\Theta \sigma$; their spinor representations read as $is_2 e^{-i\frac{\pi}{4}s_3}\tau_1\mathcal{K}$ and $s_0\tau_0\mathcal{K}$, respectively.

In the parameter regime studied in the main text (see \figref{fig:optimization}g-h), only two out of the four bands give rise to Fermi surfaces. If, furthermore, both $\alpha$ and $\Phi$ are non-zero, the two sheets do not touch, and, in the weak pairing limit \cite{DesignPrinciples}, we can describe superconductivity by a single complex number $\Delta_{\vec{k}}$ as in the first term in \equref{SCOrderParameter} for $\vec{k}$ in the vicinity of the respective Fermi surface.

Concerning quantum geometric pair breaking based on \equref{ZetaSimplified}, the limit $\alpha \gg \Phi$ corresponds to strong spin-orbit coupling, as discussed in \appref{StrongSOC}, where the spinful TRS is only weakly broken by $\Phi$ leading to a small but finite $\zeta_{\text{opt}}$. In the limit $\Phi \rightarrow 0$, the optimization problem is unfrustrated and we obtain an Anderson theorem ($\zeta_{\text{opt}}\rightarrow 0$). In the opposite limit, $\alpha \ll \Phi$, this is rather different as the problem remains frustrated: to see this, let us use the spinless TRS $\Theta$ for $\alpha\rightarrow 0$ and choose a gauge with $\ket{\phi_{-\vec{k}}}=\Theta \ket{\phi_{\vec{k}}}$. We can therefore conclude
\begin{equation}
    \tilde{I} = \sum_{\mathbf{k},\mathbf{k}'}^{\mathrm{FS}} \Delta_{\mathbf{k}}^* |W_{\mathbf{k},\mathbf{k}'}|^2 \Delta_{\vec{k}'} < \sum_{\mathbf{k},\mathbf{k}'}^{\mathrm{FS}}  |\Delta_{\mathbf{k}}|^2 |W_{\mathbf{k},\mathbf{k}'}|^2 \quad \Rightarrow \quad \zeta_{\text{opt}} > 0, \label{InequalityForItilde}
\end{equation}
where the first inequality follows from the Fermi-Dirac constraint $\Delta_{\vec{k}} = - \Delta_{-\vec{k}}$ (and continuity of $\Delta_{\vec{k}}$). In fact, \equref{InequalityForItilde} shows that quantum geometry, $|W_{\mathbf{k},\mathbf{k}'}|^2 \neq \text{const.}$, here enhances the stability of superconductivity: for $|W_{\mathbf{k},\mathbf{k}'}| = W_0$, we directly see that $\tilde{I} = 0$ and $\zeta_{\text{opt}}=1/2$ (irrespective of the form of the pairing state). This is also related to the reason why we used a full two-sublattice (with spin, four-band) model, such as \equref{AMModelNS}, instead of an effective two-band model (only with spin). A simple and natural example of the latter would be $h_{\vec{k}} = s_0 \epsilon_{\vec{k}} + \Phi s_3 (k_x^2-k_y^2)$, leading to $\ket{\phi_{\vec{k}}} = \ket{\phi_{-\vec{k}}}$; this in turn yields  $W_{\mathbf{k},\mathbf{k}'} = W_{-\mathbf{k},\mathbf{k}'}$ and hence $\tilde{I} = 0$, without any quantum geometric effects. 

    We finally note that $\zeta_{\text{opt}} = 0$ in antiferromagnets since there is a spinful TRS. Of course, this does not mean that any given pairing state will be stable, see, e.g., \refcite{PhysRevLett.131.076001}, but that there is a superconductor with an Anderson theorem. We can also explicitly demonstrate this using the model in \equref{AMModelNS}, which reduces to an antiferromagnet as long as no next-nearest-neighbor terms are present ($t_2=\alpha_2=0$). Indeed, one can see in \figref{fig:am_zeta_SI}b, where we set $\alpha_2=0$ that $\zeta_{\text{opt}} \rightarrow 0$ when $t_2 \rightarrow 0$, irrespective of the value of $\Phi$.

\section{Quantum geometric pair breaking and gauge fixing for finite $\vec{q}$}

\label{app:qg_finite_q}

As anticipated in the main text, the discussion of quantum geometric pair breaking can be readily generalized to finite momentum ($\vec{q}\neq 0$) pairing. Again assuming that $\xi_{\vec{k} +\vec{q}/2} = \xi_{-\vec{k} +\vec{q}/2}$ (perfect nesting) to suppress the kinetic pair breaking effect, \equref{ZetaSimplified} is now to be substituted by
\begin{align}
    \zeta &= \frac{\sum_{\mathbf{k},\mathbf{k}'}^{\mathrm{FS}} \left| \mathcal{W}^{\vec{q}}_{\mathbf{k},\mathbf{k}'} \Delta_{\mathbf{k}'} - \Delta_{\mathbf{k}} (\mathcal{W}^{\vec{q}}_{-\mathbf{k},-\mathbf{k}'})^* \right|^2}{4 \sum_{\mathbf{k},\mathbf{k}'}^{\mathrm{FS}} |\mathcal{W}^{\vec{q}}_{\mathbf{k},\mathbf{k}'}|^2 \sum_{\mathbf{k}}^{\mathrm{FS}} |\Delta_{\mathbf{k}}|^2 }, 
\end{align}
where we defined $\mathcal{W}^{\vec{q}}_{\mathbf{k},\mathbf{k}'} := W_{\mathbf{k}+\frac{\vec{q}}{2},\mathbf{k}'+\frac{\vec{q}}{2}}$ for notational simplicity. Consequently, the functional in \equref{ItildeOptimization} yielding the optimal gauge is to be replaced by
\begin{equation}
    \tilde{I}[\bar{\alpha}_{\vec{k}}] = \sum_{\mathbf{k},\mathbf{k}'}^{\mathrm{FS}} \Delta_{\mathbf{k}}^* \mathcal{W}^{\vec{q}}_{\mathbf{k},\mathbf{k}'} \mathcal{W}^{\vec{q}}_{-\mathbf{k},-\mathbf{k}'} \Delta_{\vec{k}'} e^{i(\bar{\alpha}_{\mathbf{k}'} - \bar{\alpha}_{\mathbf{k}})}, \label{ITildeFiniteq}
\end{equation}
where $\bar{\alpha}_{\mathbf{k}}$ are still constrained to be even in $\vec{k}$. For $\vec{q}\neq 0$, the real-space superconducting order parameter now depends on the center-of-mass coordinate $\vec{r}$ of the electrons,
\begin{equation}
    \sum_{\mathbf{k}} \Delta_{\mathbf{k}} d^{\dagger}_{\mathbf{k} + \frac{\vec{q}}{2}} d^{\dagger}_{-\mathbf{k}+ \frac{\vec{q}}{2}} = \sum_{\mathbf{r},\mathbf{x},\alpha,\alpha'} c^\dagger_{\mathbf{r}-\frac{\mathbf{x}}{2},\alpha} D_{\alpha,\alpha'}(\mathbf{x}) c^\dagger_{\mathbf{r}+\frac{\mathbf{x}}{2},\alpha'} e^{-i\vec{q}\vec{r}}.
\end{equation}
Meanwhile, the form of 
\begin{align}
    D_{\alpha,\alpha'}(\vec{x}) = \sum_{\vec{k}} \Delta_{\vec{k}} \phi_{\vec{k}+\frac{\vec{q}}{2},\alpha} \phi_{-\vec{k}+\frac{\vec{q}}{2},\alpha'} e^{i\mathbf{kx}} 
\end{align}
still closely parallels that of \equref{eq:Dx_def}. Importantly, it also still holds $C_{w}(\mathbf{x} = 0) = \tilde{I}$ with $C_{w}$ given in \equref{eq:Cx_def} and $\tilde{I}$ in \equref{ITildeFiniteq}. Consequently, the gauge optimization according to $\zeta$ is equivalent to finding the most localized Cooper pair wavefunction, even when $\vec{q}\neq 0$.

\end{document}